\documentclass[a4paper,11pt]{article}
\pdfoutput=1 

\usepackage{jcappub} 

\usepackage{float}

\usepackage[T1]{fontenc} 
\usepackage{mathtools,euscript}  

\usepackage[font=scriptsize]{caption}
\usepackage{mwe}
\usepackage{array,,multirow}
\usepackage{caption}   
\usepackage{subcaption}
\usepackage{graphicx,xcolor}
\usepackage{ragged2e}

\title{Investigating non-Gaussianity in Cosmic Microwave Background Temperature Maps using Spherical Harmonic Phases}

\author[a,1]{Sarvesh Kumar Yadav,\note{Corresponding author.}}
\author[a]{Rajib Saha}

\affiliation[a]{Indian Institute of Science Education and Research,\\Bhopal, Madhya Pradesh, India}

\emailAdd{sarveshk@iiserb.ac.in}
\emailAdd{rajib@iiserb.ac.in}

\abstract{ In the era of precision cosmology, accurate estimation of cosmological parameters is based upon the implicit assumption of the Gaussian nature of Cosmic Microwave Background (CMB) radiation. Therefore, an important scientific question to ask is whether the observed  CMB map is consistent with Gaussian prediction. In this work, we extend previous studies based on CMB spherical harmonic phases (SHP) to examine the validity of the hypothesis that the temperature field of the CMB is consistent with a Gaussian random field (GRF). The null hypothesis is that the corresponding CMB SHP are independent and identically distributed in
	terms of a uniform distribution in the interval [0, 2$\pi$] \citep{1986ApJ...304...15B,2013rossmanith}. We devise a new model-independent method where we use ordered and non-parametric Rao's statistic, based on sample arc-lengths to comprehensively test uniformity and independence of SHP for a given $\ell$ mode and independence of nearby $\ell$ mode SHP. We performed our analysis on the scales limited by spherical harmonic modes $\le$ 128, to restrict ourselves to signal-dominated regions. To find the non-uniform or dependent sets of SHP, we calculate the statistic for the data and 10000 Monte Carlo simulated uniformly random sets of SHP and use 0.05 and 0.001 $\alpha$ levels to distinguish between statistically significant and highly significant detections. We first establish the performance of our method using simulated Gaussian, non-Gaussian CMB temperature maps, along with observed non-Gaussian 100 and 143 GHz Planck channel maps. We find that our method, performs efficiently and accurately in detecting phase correlations generated in all of the non-Gaussian simulations and observed foreground contaminated 100 and 143 GHz Planck channel temperature maps. We apply our method on  Planck satellite mission's final released CMB temperature anisotropy maps- COMMANDER, SMICA, NILC, and SEVEM along with WMAP 9 year released ILC  map. We report that SHP corresponding to some of the $m$-modes are non-uniform, some of the $\ell$ mode SHP and neighboring mode pair SHP are correlated in cleaned CMB maps. The detection of non-uniformity or correlation in the SHP indicates the presence of non-Gaussian signals in the foreground minimized CMB maps. 
}
\begin{document}
\maketitle
\flushbottom

\section{Introduction}
\label{sec:intro}
One of the central ideas in many cosmological models is that galaxies and large-scale structures
in the universe have grown from small initial perturbations via a process of gravitational
instability. In one of the most successful of such models, the primordial perturbations that
seeded the initial gravitational instability were generated during a period of rapid expansion,
known as inflation \cite{1981PhRvD..23..347G,1982PhLB..108..389L,1982PhLB..117..175S}.  The perturbations produced by inflation are said to be a
statistically homogeneous Gaussian random field \cite{1986ApJ...304...15B}. It is believed that the imprint of such
perturbations are on the last scattering surface (LSS) of CMB
at large angular scales.  The statistical properties of the fluctuations on the LSS will be
highly correlated to the primordial perturbation, making it a very useful probe for testing
the Gaussianity of the primordial universe. Test of Gaussianity becomes essential from the
fact that the temperature and polarization power spectrum used to derive the cosmological
parameters, assume that the statistical properties of the primordial CMB signal is Gaussian. Their detection and identification allows to distinguish various inflationary models  \cite{2014A&A...571A..24P} making the investigation further relevant.

Unlike Gaussianity, non-Gaussianity can be of numerous types, making it difficult to detect and quantify. As detection of non-Gaussianity has far-reaching consequences for our understanding of the primordial universe \cite{2018arXiv180706211P}, hence one needs to test it with various statistical measures, each sensitive to distinct forms of the non-Gaussianity present in CMB data. Though detected non-Gaussianity might not have a primordial origin, they can still lead us to better understand the foreground residuals and systematics present in the cleaned CMB maps.

Numerous studies on non-Gaussianity for CMB missions such as COBE, MAXIMA, BOOMERanG, WMAP and Planck has been done. Most of these works are based on measures such as  bispectrum \cite{2000ApJ...528L..57M,2003NewAR..47..815H,2003MNRAS.341..623S,2010MNRAS.405..961C,2012JCAP...12..032F,2014A&A...571A..24P}, trispectrum \cite{ 2001ApJ...563L..99K,2003MNRAS.343..284D,2010arXiv1012.6039F,2010arXiv1001.5026S}, skewness and kurtosis \cite{2000ApJ...534...25C,2009PhRvD..79f3528B, 2010PhRvD..81f3533B, 2015A&A...573A.114B}, spherical Mexican hat wavelet \cite{2000MNRAS.318..475B,2004ApJ...609...22V,2004ApJ...613...51M,2005MNRAS.358..684C,2005MNRAS.359.1583M,
	2005ApJ...633..542L,2005MNRAS.356...29C,2006MNRAS.369.1858M,2006MNRAS.371L..50M,2009ApJ...706..399C,2009MNRAS.393..615C,2012MNRAS.426.1361C,2011MNRAS.417..488C}, minkowski functionals \cite{2001PhRvL..87y1303W,2002ApJ...572L..27P,2003NewAR..47..797K,2016MNRAS.461.1363N,2017CQGra..34i4002B}, directional spherical real morlet wavelet analysis \cite{2005MNRAS.359.1583M}, scaling index method \cite{2009MNRAS.399.1921R,2011AdAst2011E..11R}, method based on the N-point probability distribution function \cite{2009MNRAS.397..837V}, skeleton statistics \cite{2004ApJ...612...64E,2010MNRAS.407.2141H}, spectral distortions \cite{2017JCAP...09..042R}, neural-network \cite{2015JCAP...09..064N}, multipole vector \cite{2004PhRvD..70d3515C}, genus shift parameters \cite{2001ApJ...556..582P,2004MNRAS.349..313P}, bipolar spherical harmonics \cite{2012PhRvD..85b3010B}. In many of these studies, the estimator is based on some phenomenological model and is capable of detecting certain types of non-Gaussianity. Another way to check Gaussianity is based on a blind test, one which is motivated by statistics rather than phenomenology; this has been done in some of the above-mentioned works. Most of the above-cited works use $a_{\ell m}$ or estimators derived using them. However not many studies using phases corresponding to the complex quantity $a_{\ell m}$, which have a circular uniform random distribution, given $a_{\ell m}$ are from Gaussian random distribution, undertaken. Some of such studies are using Kuiper statistics \cite{2004MNRAS.350..989C,2006IJMPD..15.1283C,2013AN....334.1020K} , temperature-weighted extrema correlation functions \cite{2005astro.ph..5046L}, using phase mapping technique \cite{2003ApJ...590L..65C,2004ApJ...602L...1C,2004MNRAS.349..695N}, trigonometric moments of phases and Pearson's random walks  \cite{2004astro.ph..1622W,2005PhRvD..72f3512N}. The phases are related to the spherical harmonic coefficients $a_{\ell m}$ in a nonlinear way and are therefore sensitive to mode correlations of CMB in a different way. Some of the works mentioned above have reported the detection of non-Gaussianity, and others had not detected any significant signal.

In the present article, we perform a model-independent study to investigate non-Gaussianity by testing the random phase hypothesis \citep{1986ApJ...304...15B,2013rossmanith} using the Planck satellite mission's final released cleaned
CMB full-sky temperature anisotropy maps - COMMANDER, SMICA, NILC, SEVEM  along with the final release of WMAP internal linear combination map. We first apply our method to simulated Gaussian and non-Gaussian CMB temperature maps to test its performance on maps with known statistical properties. We then used observed 100 and 143 GHz Planck channel maps to test our method's performance on maps whose exact statistical properties are unknown but are known to be non-Gaussian. We use 
a new circular statistic known as the Rao's spacing test~\cite{srao}, which is sensitive to multi-mode type non-uniformities in data. We have used this statistic to test for uniformity and independence of set of SHP, absence of any one of them indicate non-Gaussianities in corresponding $a_{l m}$.  This feature of non-uniformity and dependence has not been investigated in the literature earlier. We perform
two tests to investigate uniformity in SHP of each $\ell$ and $m$ mode and four other tests to examine independence of a given $\ell$ mode and nearby $\ell$ modes SHP.

\vspace{-0.15mm}
We organize our paper as follows. In Section~\ref{formalism} we illustrate the basic formalism of this work  by first describing 
the phases derived from the  spherical harmonic coefficients and  then  presenting the details of the statistics used in the current work. 
In Section~\ref{method}, we elaborate on the methodology applied to detect the statistically significant signal in various maps. In Section~\ref{simulations}, we present a method to generate Gaussian and non-Gaussian maps, discuss statistical properties of the non-Gaussian maps and test performance of our method on various simulations. In Section~\ref{result},
we present the results. Finally, in Section~\ref{Diss}, we discuss and conclude. %

\section{Formalism}
\label{formalism}
\label{formalism}
\subsection{CMB Spherical Harmonics Phases}

CMB fluctuations, $\Delta T(\theta,\phi)$, over the last scattering surface can be mathematically expressed as, 
\begin{eqnarray}
\Delta T(\theta,\phi) = \frac{(T(\theta,\phi)-\bar{T})}{\bar{T}} = \sum_{\ell=1}^{\infty}\sum_{m=-\ell}^{\ell}a_{\ell m}Y_{\ell m}(\theta,\phi), 
\label{eqn:fluctu}
\end{eqnarray}
where $\bar{T}$ is the mean temperature over the whole sky, $T(\theta,\phi)$ is the temperature in the direction $(\theta,\phi)$ on the celestial sphere in some coordinate system. The $Y_{\ell m}(\theta,\phi)$ are spherical harmonic functions, defined in terms of the Legendre polynomials $P_{\ell m}$ as,
\begin{eqnarray}
Y_{\ell m}(\theta,\phi) = (-1)^{m}\sqrt{\frac{(2\ell+1)(\ell-m)!}{4\pi (\ell+m)! }}P_{\ell m}(\cos\theta)\exp(im\phi), 
\label{eqn:ylm}
\end{eqnarray}
where we have used the Condon-Shortley phase convention.  In equation (\ref{eqn:fluctu}), the $a_{\ell,m}$ are complex coefficients and so one can write it as,
\begin{eqnarray}
a_{\ell m}=|a_{\ell m}|\exp[i\psi_{\ell m}],
\label{eqn:alm}
\end{eqnarray}
where $\psi_{\ell m}$ is the phase angle. If $\mathcal{R}(a_{\ell m})$ and $\mathcal{I}(a_{\ell m})$ represents real and imaginary parts respectively of the complex $a_{\ell m}$ then $\Psi_{\ell m}$ is written as,
\begin{eqnarray}
\psi_{\ell m}= \frac{\mathcal{I}(a_{\ell m})}{\mathcal{R}(a_{\ell m})}.
\label{eqn:phaseS}
\end{eqnarray}

As $\Delta$ is a real quantity, $\mathcal{R}(a_{\ell m})=-\mathcal{R}({a_{\ell -m}})$ and $\mathcal{I}(a_{\ell m})=\mathcal{I}(a_{\ell -m})$ for odd $m$, $\mathcal{R}(a_{\ell m})=\mathcal{R}({a_{\ell -m}})$ and $\mathcal{I}(a_{\ell m})=-\mathcal{I}(a_{\ell -m})$ for even $m$ and $\mathcal{I}(a_{\ell m})=0$, for $m$=0. This implies that we have only $\ell$ numbers of independent phases corresponding to a given  $\ell$ mode.

\subsection{Rao's Spacing Test}

If the CMB temperature anisotropies constitute a GRF, the real and imaginary part of the $a_{\ell m}$ are both independent Gaussian distributed or equivalently, the $|a_{\ell m}|$ are Rayleigh distributed and phases $\psi_{\ell m}$ are uniformly random in $[0, 2\pi]$. The hypothesis that the phases $\psi_{\ell m}$ of the spherical harmonic coefficients $a_{\ell m}$ are independent and identically distributed in terms of a uniform distribution in the interval $(0, 2\pi]$ is known as the random phase hypothesis \cite{2013rossmanith,1986ApJ...304...15B}.This is the null hypothesis for the current work. Validity of null
	hypothesis  indicates that the  corresponding spherical harmonic coefficients are consistent with
	Gaussian random variables. If null hypothesis is violated these coefficients become
	consistent with non Gaussian variables. To test for this null hypothesis one needs to test both for the uniformity as well as independence in the SHP. In this work, we have employed Rao's \cite{raoarc} statistic to test both uniformity and independence in SHP. The method to test uniformity and independence in SHP is elaborated in the next subsection. 

Rao's test is a powerful statistic for testing uniformity of circular data. It is more powerful than the popular Rayleigh test and Kuiper's test \cite{kuiper} when the underlying circular distribution is multimodal. Most of the earlier studies on the testing of random phase hypothesis of CMB SHP have employed Kuiper's \cite{kuiper,2007ApJ...664....8C,2004MNRAS.350..989C} V statistics. Unlike Kuiper's statistics which is based on empirical distribution function, Rao's statistic is goodness$-$of$-$fits test based on the idea that if the underlying circular distribution is uniform, successive observations should be approximately evenly spaced, about $2\pi/n$ apart, where n is the number of circular samples. Large deviations from this distribution, resulting from unusually large spaces or unusually short spaces between observations, signify non-uniformity in the set of circular variables. Further, it is invariant under a choice of the origin and is non-parametric ordered statistics applicable to circular data. Suppose we have a set  of $n$ circular variables $\left\lbrace\theta_{i}\right\rbrace$, $ i=1,...n $ in the interval $ [0,2\pi] $ which are arranged in ascending order w.r.t a given zero direction and the sense of rotation, then the statistics is defined as,
\begin{eqnarray}
U_{n}=\sum_{i=1}^{n} max\left( \left\lbrace D_{i}-2\pi/n, 0 \right\rbrace \right) 
\label{eqn:un}
\end{eqnarray}
where
\begin{center}
	
	$D_{i}=\theta_{i+1}-\theta_{i}$,~~~~~for 1$\leq$i$\leq$n-1\\
	$D_{i}= 2\pi-\theta_{n}+\theta_{1}$,~~~~~~for i$=$n 
	
\end{center}

If all the circular variables are equally spaced (perfectly uniform) then $U_{n}=0$, else $U_{n}>0$, which is true for samples of finite size. Large values of $U_{n}$ indicates clustering of the variables and their distribution non-uniform.

\section{Method}
\label{method} 

In the present work we use the Rao's statistic to test for uniformity and independence of set of SHP for a given $\ell$, $m$ modes and independence of SHP for $\ell$-mode pair. For an independent set of phases, the phase differences must be circular uniform distributed in $(0,2\pi]$\cite{2007ApJ...664....8C,2004MNRAS.350..989C}. In order to test for uniformity of these SHP differences we use the Rao's statistic. Detection of non-uniformity in SHP differences indicates presence of correlation among the set of SHP in question. In this article we apply our method on phases obtained from full-sky simulated and data maps. We first use simulated Gaussian and non-Gaussian maps to test performance of our method on maps with known statistical properties. Second, we apply the method on phases of the full-sky foreground-contaminated Planck satellite mission's 100 GHz and 143 GHz frequency bands. Due to the presence of foreground, these maps can be considered as observed non-Gaussian maps. Finally, we apply our method on COMMANDER, SMICA, NILC, SEVEM \cite{2018arXiv180706208P} temperature maps from the latest Planck release, and WMAP-ILC \cite{2013ApJS..208...20B} temperature map from the final WMAP release. As all the above cleaned CMB maps correspond to the same CMB realization, any cosmological signal must be consistently detected in all or at least most of them once its origin from any foreground or systematic is ruled out. Usually to improve the signal to noise ratio we convolve a CMB map with some Gaussian filter.  Though this modifies the $a_{\ell m}$ , they do not have any effect on the SHP. The map $a_{\ell m}$  for a given $\ell$ and $m$ can be written as,


\begin{eqnarray*}
	a_{\ell m} = a_{\ell m}^S + a_{\ell m}^N
\end{eqnarray*}
where $a_{\ell m}^S$ and $a_{\ell m}^N$ represents signal and noise components respectively. Hence the phases of the map corresponding to a given $\ell$ and $m$ can be written as, 
\begin{eqnarray*}
	\psi_{\ell m} = \frac{\mathcal{I}(a_{\ell m}^S)sin(\psi_{lm}^S) + \mathcal{I}(a_{\ell m}^N)sin(\psi_{lm}^N)}{\mathcal{R}(a_{\ell m}^S)cos(\psi_{lm}^S) + \mathcal{R}(a_{\ell m}^N)cos(\psi_{lm}^N)}
\end{eqnarray*}
where $\mathcal{R}$ represent real part and $\mathcal{I}$ depict imaginary part. The
$\psi_{lm}^S$ and $\psi_{lm}^N$ are the signals and the noise phase, respectively.
 From the above phase expression, it is clear that the $\psi_{\ell m}$ will be dominated by signal 
 iff $a_{\ell m}^S$ >> $a_{\ell m}^N$. 
 For CMB temperature maps, this condition holds for low $\ell$, and hence in order to avoid large 
 contribution to SHP from noise, we restrict our analysis to a maximum $\ell$ mode of 
 128. 
 
  We employ ianafast facility in Healpix package to 
 obtain the spherical harmonic coefficient i.e., $a_{\ell m}$ for each of the above maps. Using the above $a_{\ell m}$  we obtain 
$\psi_{\ell m}$ applying (\ref{eqn:phaseS}) for $\ell$ mode 4 to maximum $\ell$ mode 128 for all 
the maps. From the above-acquired SHP, we investigate them for uniformity and independence. 

We divide our analysis in three classes. In the first class (class I), we investigate for uniformity of SHP for constant $\ell$ and $m$ modes. In the second class (class II) we examine correlation within SHP for a given $\ell$ mode. And finally, in the third class (class III) we investigate independence of SHP of neighbouring $\ell$ modes. We have following two variable types for class I :
\justify{
	\textbf{Type} $(i)$ : $\left\lbrace \psi_{\ell=constant,m}^i \right\rbrace$, set of all phases for a given $\ell$ mode, for testing the uniformity of phases in each $\ell$-mode.\\
	\textbf{Type} $(ii)$ : $\left\lbrace \psi_{\ell,m=constant}^i \right\rbrace$, set of all phases for a given $m$ mode, for testing the uniformity of phases with given $m$-mode and all $\ell$-modes.\\
}

\hspace{-1.0cm} For the class II, we use two types of variables represented by following in our analysis :
\justify{
	\textbf{Type} $(iii)$ : $\left\lbrace \psi_{\ell=constant,m+1}^i - \psi_{\ell=constant,m}^i \right\rbrace$, difference of phases for a given $\ell$ but consecutive $m$, to test the correlation between consecutive $m$ modes.\\
	\textbf{Type} $(iv)$ : $\left\lbrace \psi_{\ell=constant,m+2}^i - \psi_{\ell=constant,m}^i \right\rbrace$, difference of phases for a given $\ell$ but next to consecutive $m$, to test the correlation between next to consecutive $m$ modes.\\
}

\hspace{-1.0cm} For the class III, we use two types of variables represented by following in our analysis :
\justify{
	
	\textbf{Type} $(v)$ : $\left\lbrace \psi_{\ell+1,m}^i - \psi_{\ell,m}^i \right\rbrace$, difference of phases of two consecutive $\ell$ having same $m$, to test the correlation between consecutive $\ell$ modes but same $m$. For example if $\psi_{\ell,m}$ denotes $m^{th}$ phase of multipole $\ell$, the phase differences of multipole pair $(6,7)$ is the following set of variables :
	$[\psi_{7,1}$ - $\psi_{6,1}$, $\psi_{7,2}$ - $\psi_{6,2}$, $\psi_{7,3}$ - $\psi_{6,3}$, $\psi_{7,4}$ - $\psi_{6,4}$, $\psi_{7,5}$ - $\psi_{6,5}$, $\psi_{7,6}$ - $\psi_{6,6}]$.\\
	\textbf{Type} $(vi)$ : $\left\lbrace \psi_{\ell+2,m}^i - \psi_{\ell,m}^i \right\rbrace$, difference of phases of next to consecutive $\ell$ having same $m$, to test the correlation between next to consecutive $\ell$ modes but same $m$. For example if $\psi_{\ell,m}$ denotes $m^{th}$ phase of multipole $\ell$, the phase differences of multipole pair $(6,8)$ is the following set of variables :
	$[\psi_{8,1}$ - $\psi_{6,1}$, $\psi_{8,2}$ - $\psi_{6,2}$, $\psi_{8,3}$ - $\psi_{6,3}$, $\psi_{8,4}$ - $\psi_{6,4}$, $\psi_{8,5}$ - $\psi_{6,5}$, $\psi_{8,6}$ - $\psi_{6,6}]$.
	
}
\hspace{.5cm} 

The analysis is performed on all the above variable after bringing them in range $(0,2\pi]$. As all the above variables are expected to be uniformly distributed in $(0,2\pi]$, we use the Rao statistic (\ref{eqn:un}) to get values of $U_{n}$ for all of the above six cases and each map. To examine whether the obtained $U_{n}$ value is anomalous (corresponds to non-uniform distribution), we compare it against $U_{n}$ obtained from 10000 Monte Carlo simulated sets of random uniform circular variables in $(0,2\pi]$ with number of phases in each set depending on number of phases involved in a given $\ell$ and $m$ analysis. For a given set of SHP or difference of SHP, we arrange the $U_{n}$ values obtained from 10000 Monte Carlo simulation in ascending order and consider any value of $U_{n}$ obtained from data greater than or equal to 95$\%$ of simulations to be significant. We use this idea as we expect large value of $U_{n}$ is associated with non-uniformity in a given sample. We plot U versus $\ell$ or $m$ modes for all different maps together for each case in Figure \ref{fig:sim-t1} to Figure \ref{fig:cm-6} and mark 0.950 and 0.999 p levels to ascertain significant and highly significant correlated occurrences, with 0.05 and 0.001 probability of type I errors respectively.


\section{Simulations}
\label{simulations} 

In this section, we discuss the statistical properties and the efficiency of Rao statistic when applied to non-Gaussian CMB Monte Carlo simulations. We also discuss a method to generate Gaussian and non-Gaussian CMB Monte Carlo simulations used to test the performance of our method.

\subsection{Monte Carlo Simulations of Gaussian and Non-Gaussian CMB Maps}
\label{simgen}
 We simulate Gaussian CMB map using \texttt{synfast} facility in  \texttt{HEALPix} at \texttt{Nside}=64 and \texttt{lmax}=128. To simulate non-Gaussian temperature fields on a sphere we use following steps :
 
 \justify{
 	
 	\textbf{Step 1 :} We sample random variables from normal distribution. The number of such normal deviates equals the total number of spherical harmonic coefficients for a CMB map at  \texttt{Nside}=64 and \texttt{lmax}=128. In order to generate samples from a non-Gaussian distribution, we first sample from two different regions of the normal distribution, say $-a$ to $-b$ and $b$ to $a$ where $|a|>|b|$.\\
  	\textbf{Step 2 :} Shift sampled values in $-a$ to $-b$ by $2b$ i.e.
 	\begin{eqnarray*}
	[-a,-b] \longrightarrow [-a+2b,-b+2b].
\end{eqnarray*}
 	\textbf{Step 3 :} To get the spherical harmonic coefficient $ a_{\ell m}$ from the above set of sampled variables we multiply by the beam ($B(\ell)$), pixel ($w_{pix}(\ell)$) and spectrum window ($[C(\ell)]^{1/2}$), 
 	\begin{eqnarray*}
 		x_{\ell m} \longrightarrow a_{\ell m}[C(\ell)]^{1/2}B(\ell)w_{pix}(\ell)
 	\end{eqnarray*}
    where $x_{\ell m}$ is sampled variable corresponding to $(\ell,m)$ mode. \\
    \textbf{Step 4 :} We set the imaginary part of zeroth spherical harmonic coefficients, $a_{\ell 0}$, corresponding to all $\ell$ to zero. We use the resulting $a_{\ell m}$ coefficients in \texttt{synfast} facility of \texttt{HEALPix} package to obtain the simulated non-Gaussian maps with an angular power spectrum. $C(\ell)B(\ell)^{2}w_{pix}^{2}(\ell)$.
    
    \begin{figure}[H]
    	\hspace{-.4cm}
    	\includegraphics[scale=0.54]{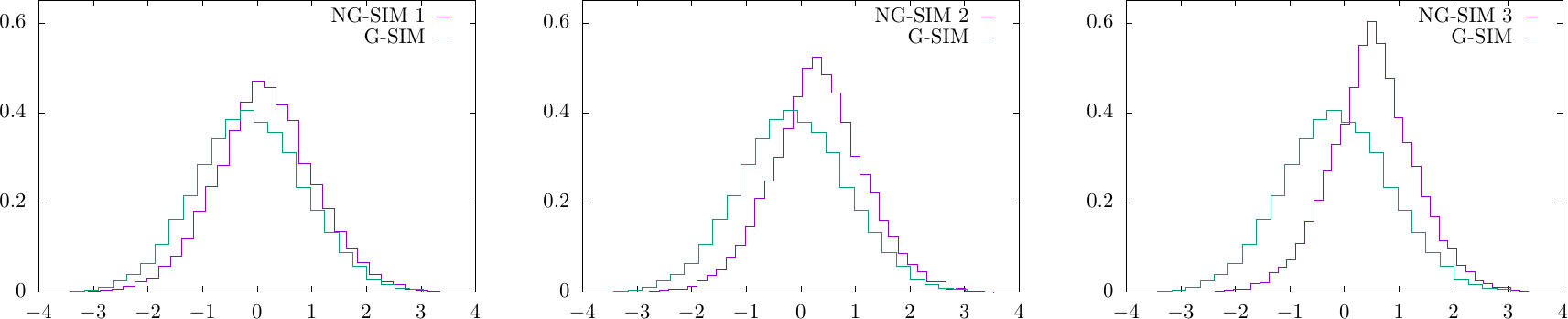}
    	\caption{In the above figure, we show density histogram for samples from different non-Gaussian Monte Carlo simulations. The leftmost figure show histogram for samples from NG-SIM 1 along with a normal curve in green. Similarly, the middle and right figure shows a histogram for samples from NG-SIM 2 and 3. This figure shows that the samples are drawn from distributions that are increasingly deviating from a normal distribution.}
    	\label{fig:histx}
    \end{figure}
 	
} 

\begin{figure}[H]
	\hspace{-.4cm}
	\includegraphics[scale=0.54]{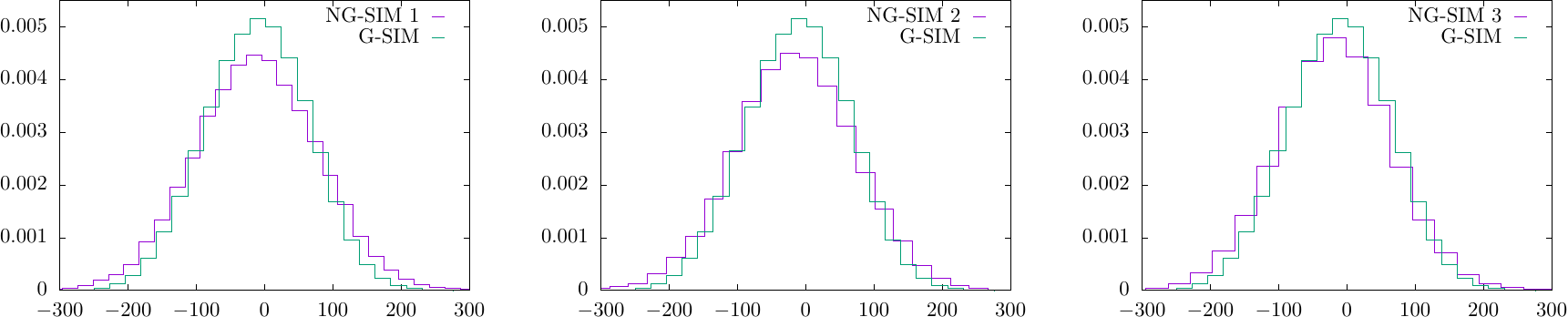}
	\caption{We show density histogram for various Gaussian and non-Gaussian Monte Carlo simulated CMB map pixel values in the above figure. We have plotted the histograms for pixel values -300 to 300 for all four maps for ease of comparison, though there is some insignificant number of outliers in the highly non-Gaussian case.	Because of central limit theorem actual signal that was present in the spherical harmonic space get obscured in the 1-point statistics in the pixel space. This warrants that special care must be taken into account to study non-Gaussianity either in spherical harmonic space or pixel space.}
	\label{fig:histpix}
\end{figure}

\begin{figure}
	\centering
	\vspace{-2.7cm}
	\begin{subfigure}{\linewidth}
		\centering+
		\includegraphics[scale=0.34]{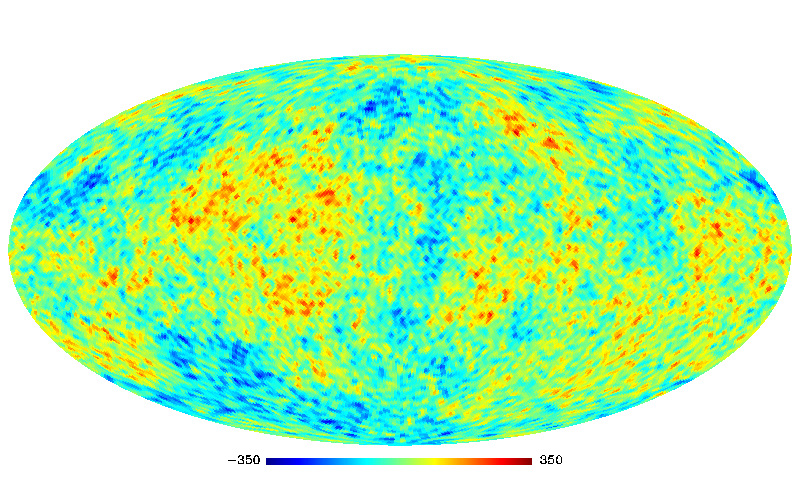}
	\end{subfigure}
	\hfill
	\begin{subfigure}{\linewidth}
		\centering
		\includegraphics[scale=0.34]{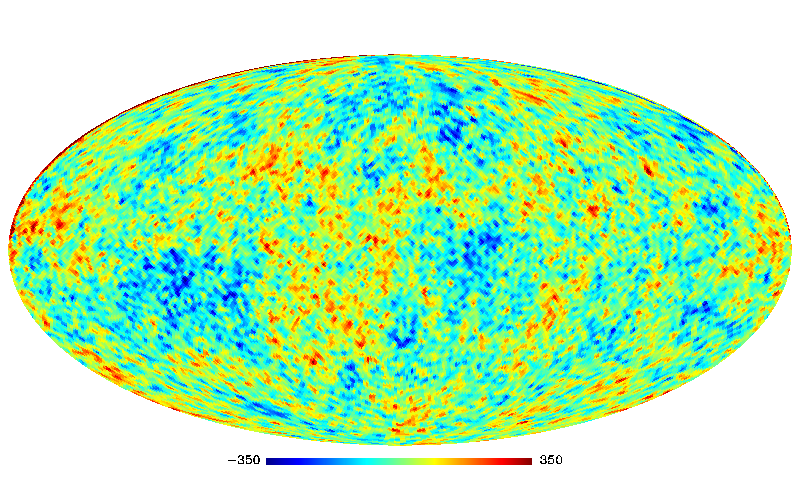}
	\end{subfigure}
	\hfill
	\begin{subfigure}{\linewidth}
		\centering
		\includegraphics[scale=0.34]{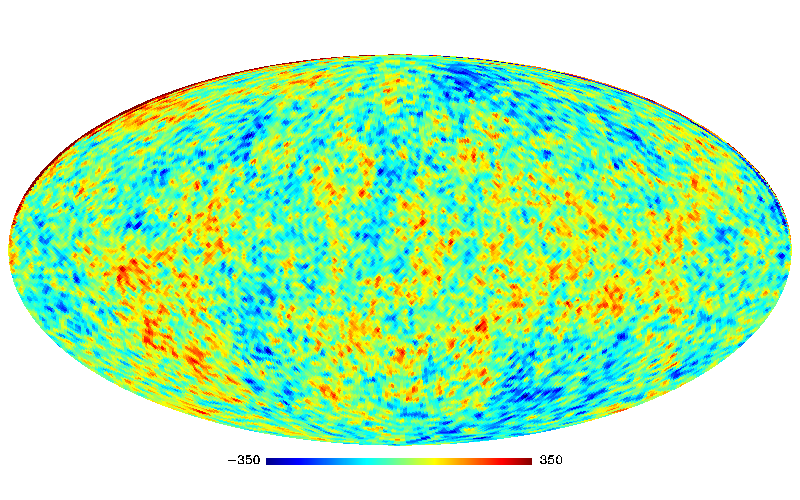}
	\end{subfigure}
	\hfill
	\begin{subfigure}{\linewidth}
		\centering
		\includegraphics[scale=0.34]{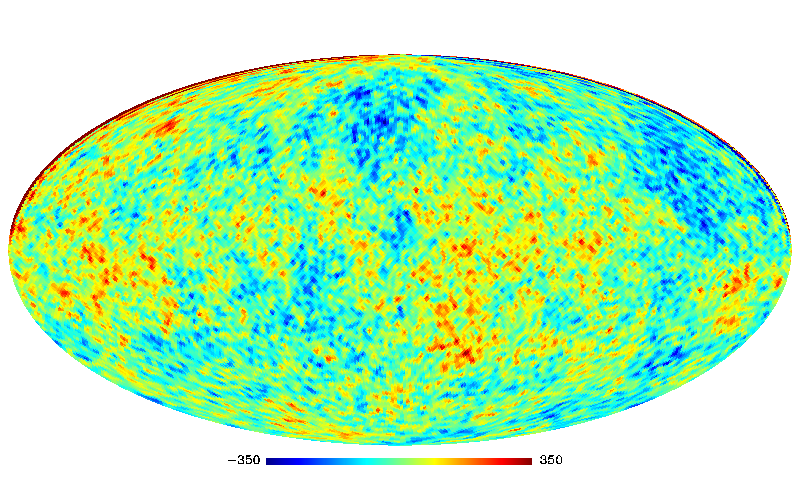}
	\end{subfigure} 
	\caption{In the above figures, we show the Mollweide projection for Gaussian and non-Gaussian Monte Carlo simulations generated using steps enumerated in section (\ref{simgen}).We simulate Gaussian CMB map using \texttt{synfast} facility in  \texttt{HEALPix} at \texttt{Nside}=64 and \texttt{lmax}=128. The topmost map corresponds to a Gaussian realization of CMB. The second, third, and fourth Mollweide projected maps from the top correspond to NG-SIM 1, 2, 3, respectively. Visually we do not see any unusual features in the three non-Gaussian maps compared to the Gaussian map in the top panel. This is expected since pixel values are linear combination of spherical harmonic coefficients, due to the central limit theorem, even for non-Gaussian spherical harmonic coefficients the resulting pixel values are random Gaussian variate.}
	\label{fig:ng}
\end{figure}

This article uses three different sets of $[a,b]$ to generate a distribution with increasing deviation from a normal distribution. The least non-Gaussian samples, which we call non-Gaussian simulation 1 (NG-SIM 1) is sampled using values with $|a|=3.50$, $|b|=0.25$, samples with higher non-Gaussianity which we call non-Gaussian simulation 2 (NG-SIM 2), is sampled using values with $|a|=3.50$, $|b|=0.40$, and finally, in order to get the most non-Gaussian samples, which we call non-Gaussian simulation 3 (NG-SIM 3), we use values with $|a|=3.50$, $|b|=0.60$. We show density histograms corresponding to samples for NG-SIM 1, 2 and 3 in Figure (\ref{fig:histx} ). We also plot density histogram corresponding to Gaussian samples in each of the sub-figures for comparison. The figure shows that the histogram for NG-SIM 1, 2, 3 deviates from a Gaussian case, with NG-SIM 1 deviating least and NG-SIM 3 the most. We show density histograms for pixel values corresponding to a single realization of the three non-Gaussian map types in the Figure (\ref{fig:histpix}) along with the histogram corresponding to a Gaussian realization in each of the sub-figures. In left sub-figure of Figure (\ref{fig:histx}) the deviation of density histogram for ``NG-SIM 1'' is minor with respect to other two non-Gaussian cases. Corresponding pixel density histogram in left sub-figure of Figure (\ref{fig:histpix}) also deviates from the gaussian case shown in green. However changes in other histograms of Figure (\ref{fig:histpix}) remain insensitive to changes in corresponding density histograms of Figure (\ref{fig:histx}). This could be explained if we look into equation (\ref{eqn:fluctu}). From the equation (\ref{eqn:fluctu}) we can see that the pixel values are superposition of spherical harmonic coefficients. Therefore even if spherical harmonic coefficients are not from Gaussian distribution, due to central limit theorem  the map pixel value tend to a Gaussian random variate. We also show the Mollweide projection for the simulated Gaussian and non-Gaussian maps in Figure (\ref{fig:ng}). The topmost map is the Gaussian realization, with increasing non-Gaussianity in spherical harmonic domain for subsequent maps. Due to the central limit theorem, visually we do not see any unusual features in the three non-Gaussian maps compared to the Gaussian map in the top panel of the Figure (\ref{fig:ng}).  

\subsection{Performance testing}
\label{simeff}

\begin{figure}[H]
		\vspace{-.2cm}
	\hspace{.5cm}
	\includegraphics[scale=0.9]{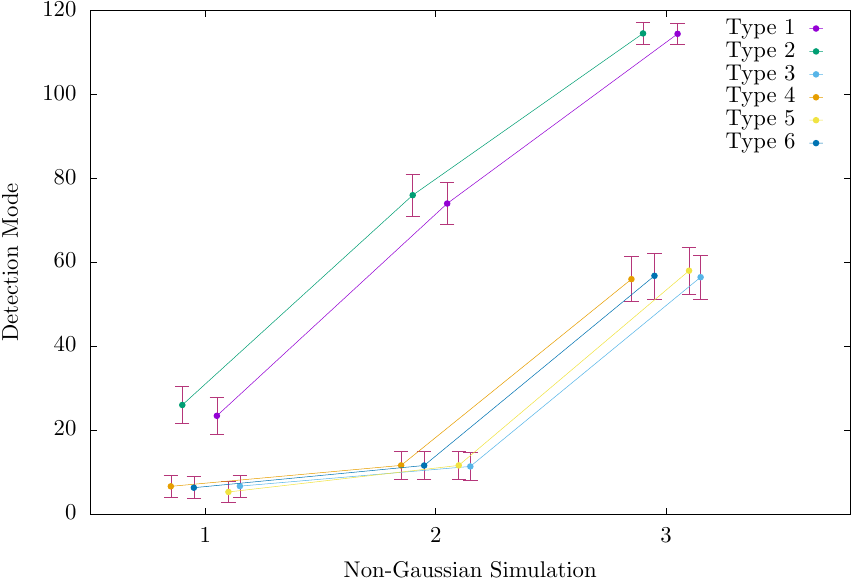}
	\caption{The above figure shows modes for all the six tests and non-Gaussian map types with corresponding error bars. The horizontal axis represents the non-Gaussian simulation set number in the plot, and the vertical axis represents the number of significant detection. Each line in the figure connects the modes corresponding to a different set of non-Gaussian maps with the same test type. To make results corresponding to test types 2, 4, 6, visible we have shifted their modes and error bars on the left along the x-axis by  0.1, 0.15, 0.05, respectively. For a similar reason, we have shifted modes and error bars on the right along the x-axis by  0.05, 0.15,  0.1 for test types 1,  3, 5, respectively. For all the six test, number of detection increase with non-Gaussianity, indicating our method performs well. The small error bar around each test and set of the non-Gaussian map indicate that our tests work efficiently in detecting significant cases.}
	\label{fig:effic}
\end{figure}

\begin{figure}[H]
	\hspace{-1cm}
	\includegraphics[scale=0.5]{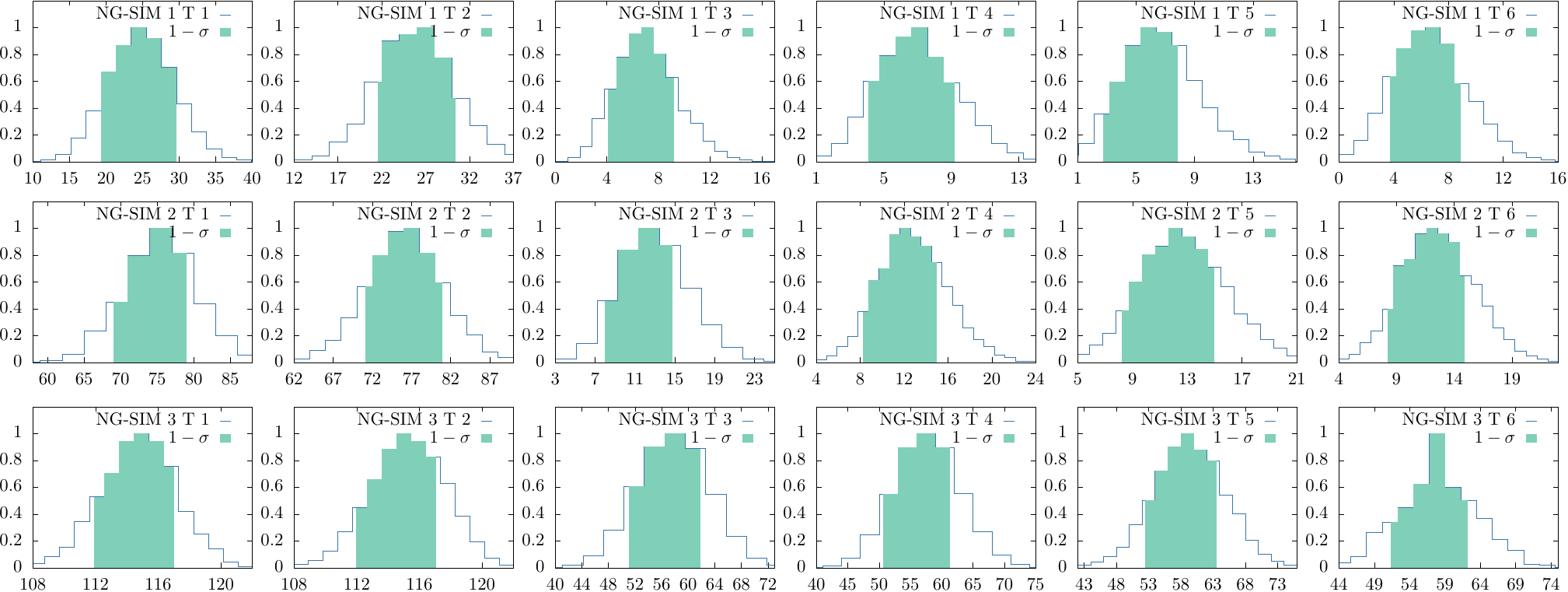}
	\caption{In the above figure, we show histograms for the number of significant detection for each type of test and all the three non-Gaussian map types. The key in Figure (\ref{fig:mltplt}) ``NG-SIM x T y'' denotes `x' type non-Gaussian simulation and test variable type `y'. This histogram plot indicates a small spread in the number of significant detection for each test type and every non-Gaussian map type indicating our method performs well. The light green region shows the 1 sigma around the mean hence are asymmetric about histogram peaks.}
	\label{fig:mltplt}
\end{figure}

We test the performance of our method using 5000 Monte Carlo simulations for all three types of non-Gaussian maps. We apply the six types of tests discussed in section (\ref{method}) on each of the Monte Carlo simulated maps. We present a histogram for the number of significant detection for each type of test and all the three non-Gaussian map types in Figure (\ref{fig:mltplt}). The key in Figure (\ref{fig:mltplt}) ``NG-SIM x T y'' denotes \textbf{`x'} type non-Gaussian simulation and test variable type \textbf{`y'}. This histogram plot indicates a small spread in the number of significant detection for each test type and every non-Gaussian map type indicating our method performs well. We also present a plot showing modes from histogram plots of Figure (\ref{fig:mltplt}) for all the six tests and the three non-Gaussian map types with corresponding error bars in the Figure (\ref{fig:effic}). In Figure (\ref{fig:effic}), the horizontal axis represents the non-Gaussian simulation set number, and the vertical axis represents the number of significant detection. Each line in the figure connects the modes corresponding to a different set of non-Gaussian maps with the same test type. To make results corresponding to test types 2, 4, 6, visible we have shifted their modes and error bars on the left along the x-axis by  0.1, 0.15, 0.05, respectively. For a similar reason, we have shifted modes and error bars on the right along the x-axis by  0.05, 0.15,  0.1 for test types 1,  3, 5, respectively. For all the six test, number of detection increase with non-Gaussianity, indicating our method performs well. The small error bar around modes for all the six tests and non-Gaussian maps indicates that our tests perform well in detecting significant cases. 

\section{Results}
\label{result}
In this section, we summaries the results we obtain after utilizing the Rao statistic on various variable types, as elaborated in section (\ref{method}), for testing uniformity of and correlations among spherical harmonic phases from various maps. In each of the Figures (\ref{fig:sim-t1}-\ref{fig:cm-6}), the red and black dashed lines represent highly significant ($\alpha=0.001$) and significant ($\alpha=0.05$) levels respectively. In the following, we present the results obtained for Monte Carlo simulated Gaussian and non-Gaussian CMB maps in subsection (\ref{result-0}). We present the results for foreground contaminated 100 and 143 GHz Planck frequency maps in subsection (\ref{result-1}). And finally, we present results obtained after applying our method on foreground cleaned CMB maps in subsection (\ref{result-2}). We apply our method only on cases with minimum phase sample size of four\footnote{ Although the minimum sample size one can use to apply the Rao's spacing test is two, with a very small sample size, the test tends to decrease the probability of detecting the alternatives of non-uniformity even when that is the case. Therefore, conservatively we choose four as the minimum sample size in our analysis. See \cite{2005MNRAS.357....1R,raoarc} also for similar lower bound on sample size.}.

\subsection{Simulated Gaussian and non-Gaussian CMB Maps}
\label{result-0} 
\begin{figure}[H]
	\hspace{-1.2cm}
	\includegraphics[scale=0.58]{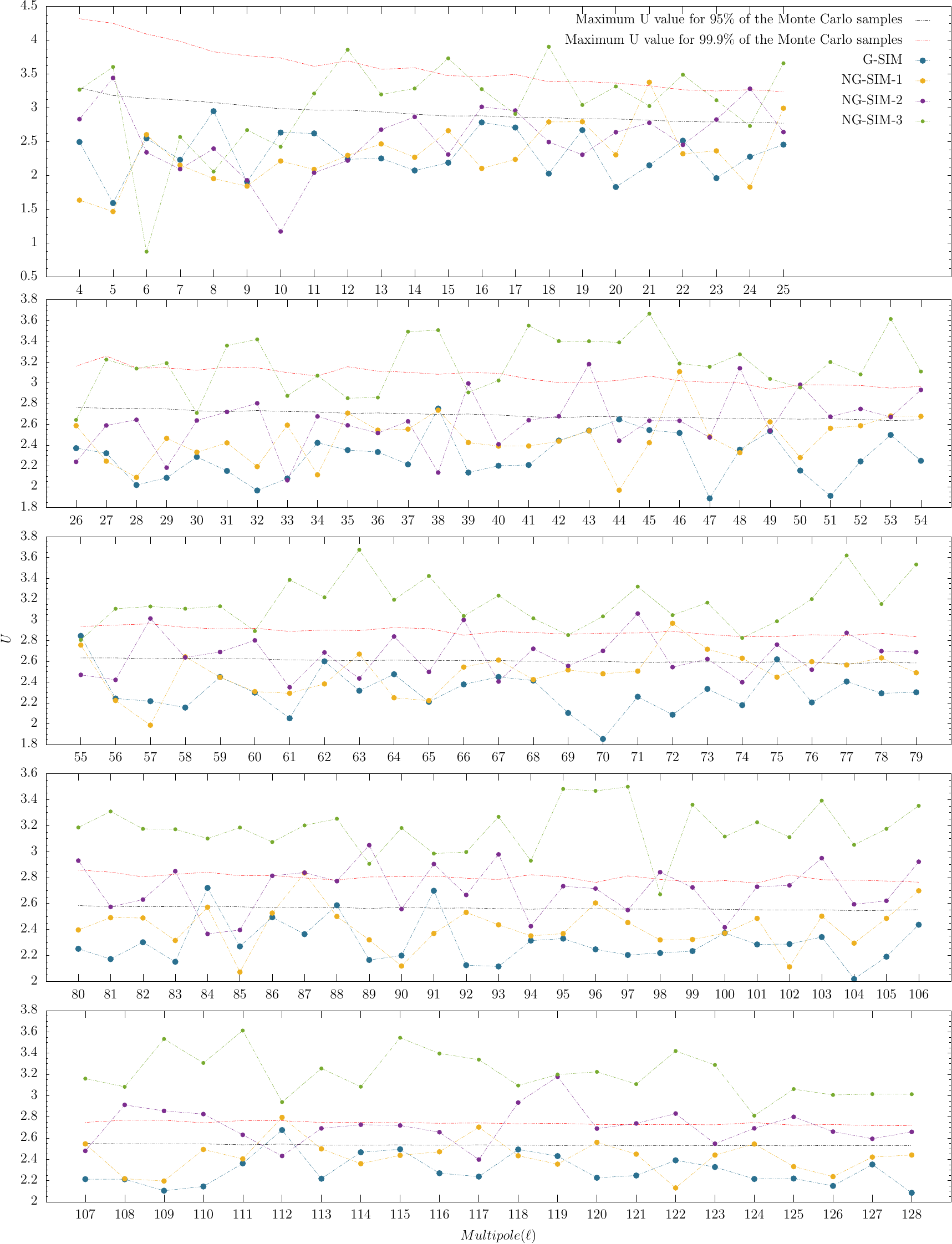}
	\caption{In the above figure we show $U$-values on vertical for testing uniformity of the given set of phases (variable of Type $(i)$) for various $\ell$-modes on horizontal axis of Monte Carlo simulated Gaussian and non-Gaussian CMB maps. We also show the maximum of $U$ for 95\% and 99.9\% of the corresponding Monte Carlo simulations.}
	\label{fig:sim-t1}
\end{figure}

\begin{figure}[H]
	\vspace{-.8cm}
	\hspace{-1.2cm}
	\includegraphics[scale=0.58]{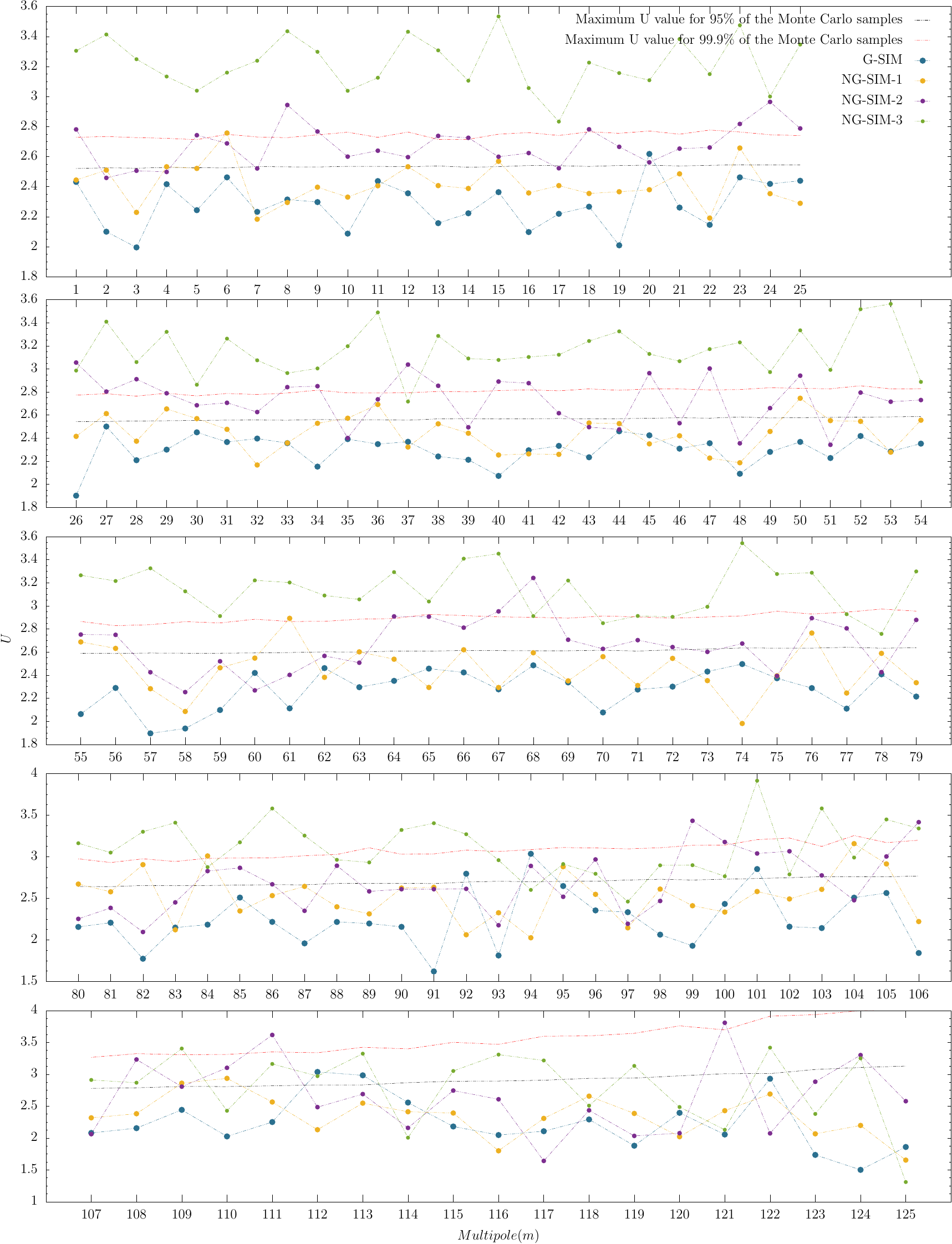}
	\caption{The figure shows the $U$-value on vertical axis and $m$-modes on the horizontal axis for testing uniformity of SHP for given $m$ and all $\ell$ (variable of Type $(ii)$) for Monte Carlo simulations of Gaussian and non-Gaussian CMB maps. We also show the maximum of $U$ for 95\% and 99.9\% of the corresponding simulated uniform phases in $(0,2\pi]$.}
	\label{fig:sim-t2}
\end{figure}

\begin{figure}[H]
	\vspace{-.8cm}
\hspace{-1.2cm}
\includegraphics[scale=0.58]{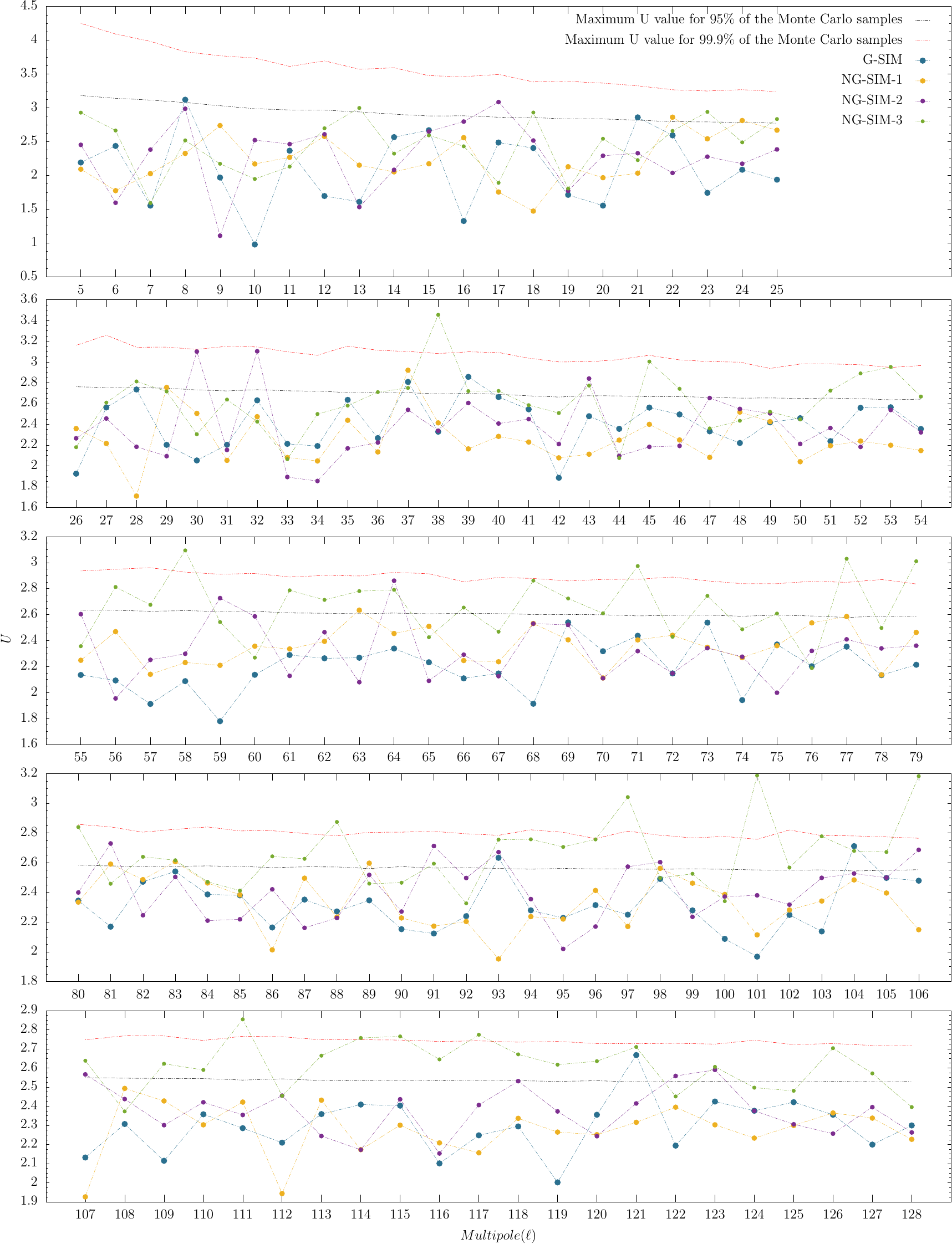}
	\caption{The above figure shows the $U$-value obtained for simulated Gaussian and non-Gaussian CMB maps using the variable of Type $(iii)$. The vertical and horizontal axes show $U$-value and $\ell$-modes, respectively. We also show the maximum of $U$ for 95\% and 99.9\% of the corresponding simulated uniform phases in $(0,2\pi]$.}
	\label{fig:sim-t3}
\end{figure}

\begin{figure}[H]
	\vspace{-.8cm}
\hspace{-1.2cm}
\includegraphics[scale=0.58]{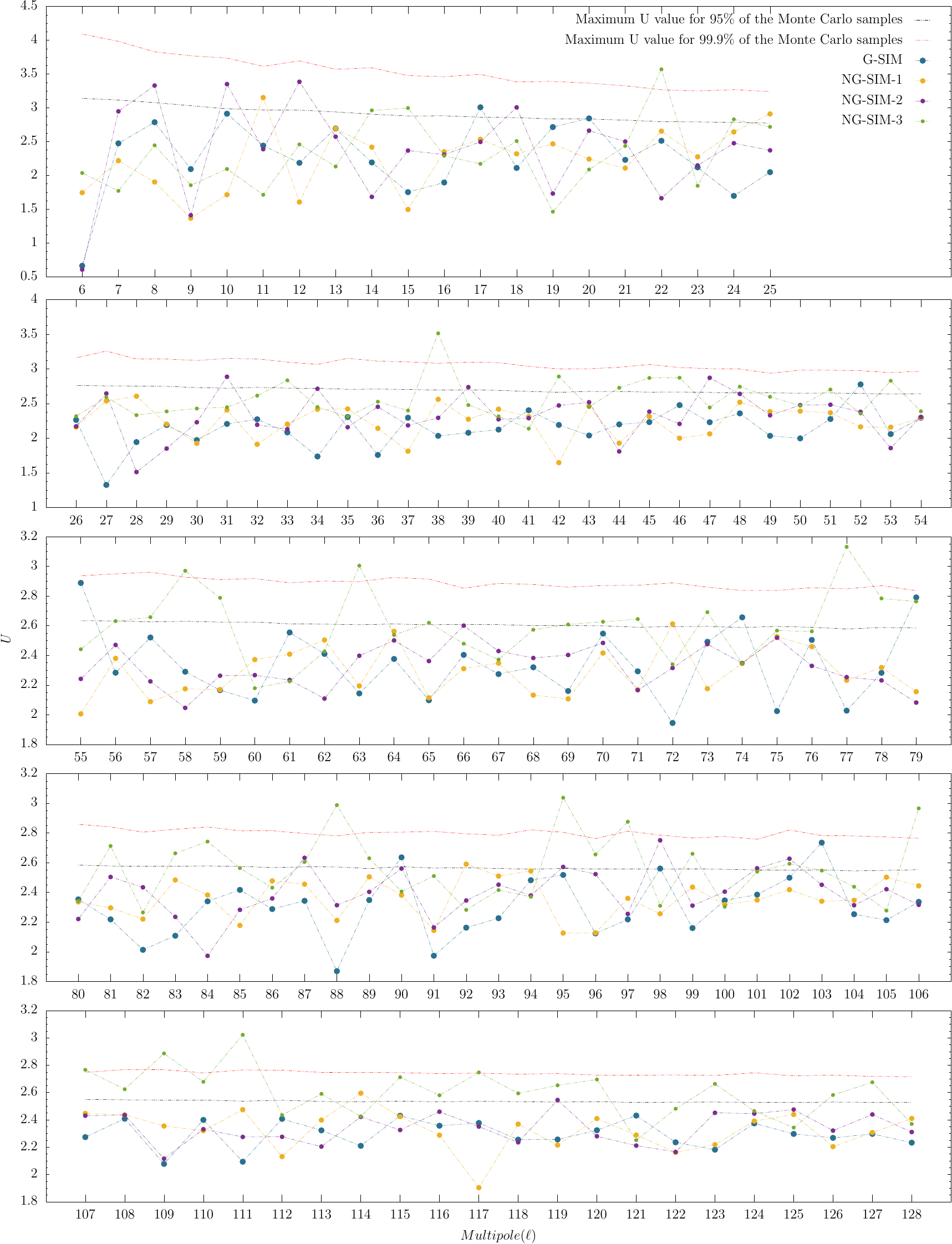}
	\caption{The above figure shows $U$-value obtained using the variable of Type $(iv)$ on phases for Gaussian and non-Gaussian simulated CMB maps. We show $\ell$-modes  on the horizontal axis and $U$-values on the vertical axis. We show the maximum $U$ for 95\% and 99.9\% of the corresponding simulated uniform phases in $(0,2\pi]$ in black and red dashed lines, respectively.}
	\label{fig:sim-t4}
\end{figure}

\begin{figure}[H]
	\vspace{-1cm}
\hspace{-1.2cm}
\includegraphics[scale=0.58]{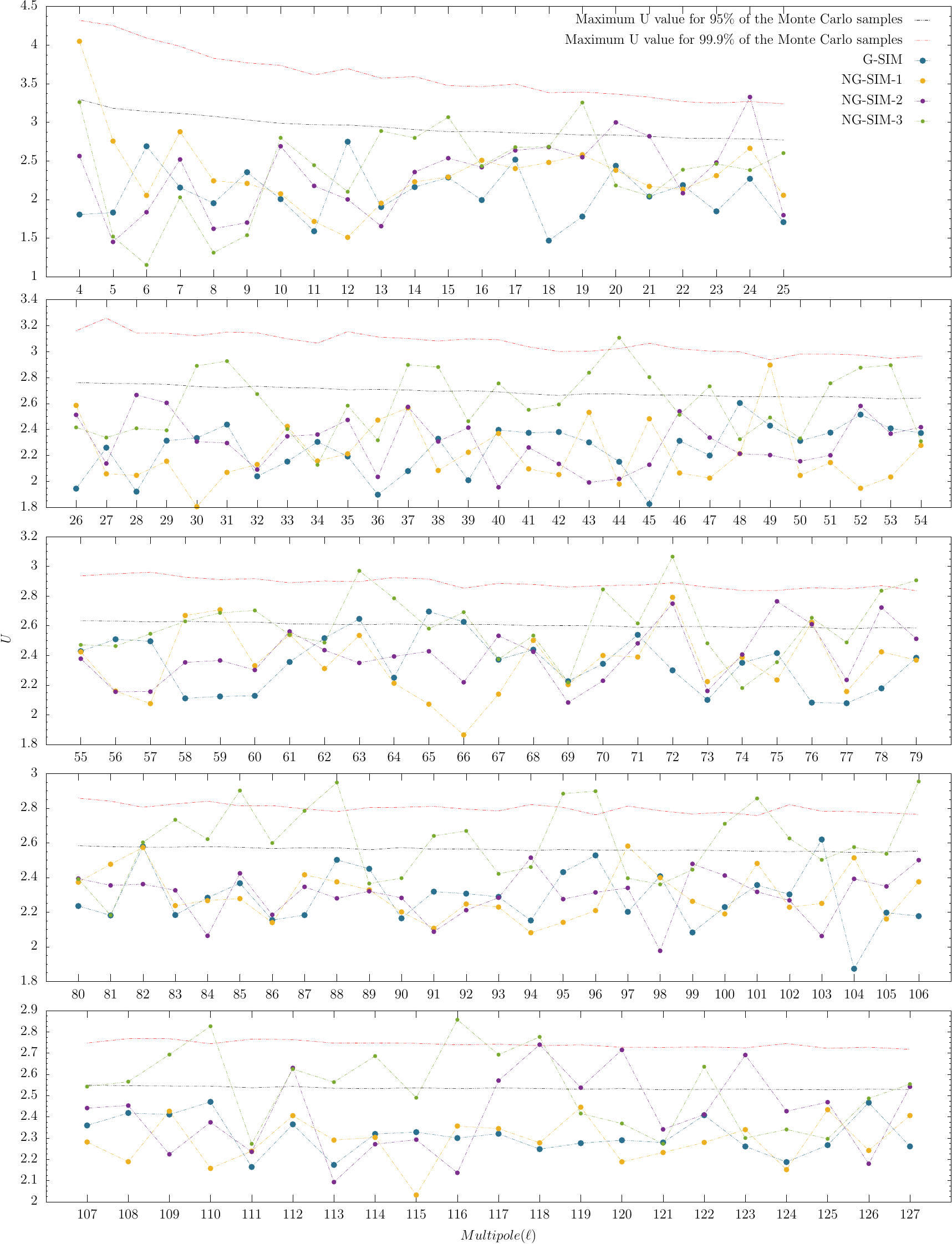}
	\caption{The figure shows $U$-value obtained using variable of Type $(v)$ for simulated Gaussian and non-Gaussian CMB maps. The vertical and horizontal axes show $U$-value and lower of the $\ell$-mode pair respectively. We also show the maximum of $U$ for 95\% and 99.9\% of the corresponding simulated uniform phase samples in $(0,2\pi]$ in black and red dashed lines respectively.}   
	\label{fig:sim-t5}
\end{figure}

\begin{figure}[H]
	\vspace{-1cm}
\hspace{-1.2cm}
\includegraphics[scale=0.58]{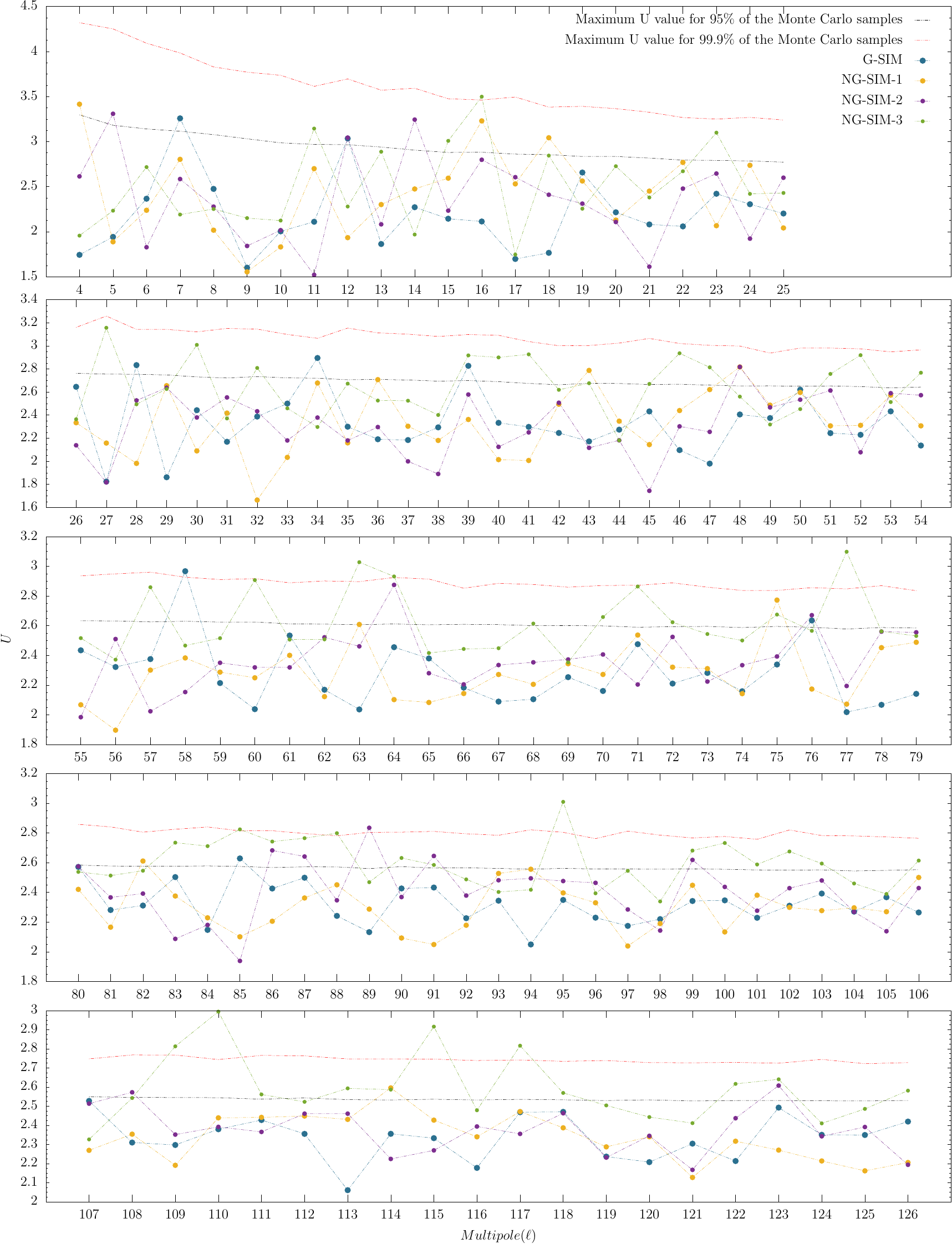}
	\caption{The figure shows $U$-value obtained using variable of Type $(vi)$ for Gaussian and non-Gaussian CMB maps. The vertical and horizontal axes show $U$-value and lower of the $\ell$-mode pair respectively. We also show the maximum of $U$ for 95\% and 99.9\% of the corresponding simulated uniform phase samples in $(0,2\pi]$ in black and red dashed lines respectively.}
	\label{fig:sim-t6}
\end{figure}

This subsection presents results obtained after applying our method on Monte Carlo simulated Gaussian and non-Gaussian maps. We apply our method to these simulations as it is essential to test any new method with simulations of known statistical properties. We present $U$-values for six tests described in section (\ref{method}) for the maps in Figure (\ref{fig:sim-t1}) to (\ref{fig:sim-t6}). The $U$-values obtained for the Gaussian CMB phases are denoted by “G-SIM” as shown in the key of Figures (\ref{fig:sim-t1}) to (\ref{fig:sim-t6}). We represent each of the Non-Gaussian simulations by “NG-SIM J” as shown in the key of Figure (\ref{fig:sim-t1}) to (\ref{fig:sim-t6}) where the value of J=1,2,3 for increasing non-Gaussian maps in the same order. 

In the Figure (\ref{fig:sim-t1}), we plot $U$-values on the y-axis obtained for Monte Carlo simulated Gaussian and non-Gaussian maps for the variable of Type $(i)$. On the plot, the x-axis represents the $\ell$-mode whose phases we test for uniformity. For the Gaussian simulated CMB map, we find seven modes are significant. For the Monte Carlo simulated non-Gaussian maps, we find 23, 69,116 significant cases corresponding to non-Gaussian cases $J=1,2,3$. In the Figure (\ref{fig:sim-t2}), we plot $U$-values on the y-axis and $m$-modes on the x-axis corresponding to the variable of Type II for all the simulated maps. We use the variable of Type $(ii)$ to test the uniformity of phases with the same $m$-mode and all the $\ell$-modes. We find six significant cases in the simulated CMB map for this case. For the non-Gaussian simulated CMB maps, we find 23, 74, 116 significant cases corresponding to non-Gaussian cases with $J= 1,2,3$.  We show $U$-values on y-axis obtained using variable of Type $(iii)$ in Figure (\ref{fig:sim-t3}) for all the simulated maps. The x-axis shows the $\ell$-modes corresponding to which phase differences are used for the test. We find seven significant cases in the Gaussian simulated map. In the non-Gaussian maps, we find 9, 15, 67 significant cases corresponding to $J=1,2,3$, respectively. In Figure (\ref{fig:sim-t4}) we show $U$-values on y-axis obtained for variable of Type $(iv)$ for all the simulated maps. In the plot, the x-axis represents the $\ell$-mode corresponding to which $m$-mode phase differences are used as test variables. For the Gaussian simulation, we find eight significant cases. For the non-Gaussian maps, we find 5, 13, 51 significant cases corresponding to $J=1,2,3$, respectively. Figure (\ref{fig:sim-t5}) shows $U$-values obtained after applying Rao's statistic on the variable of Type $(v)$ for all the simulated maps. We show $U$-values on the y-axis and lower of the $\ell$-mode pair involved in the analysis on the x-axis. For example, for $\ell$-mode pair $(6,7)$, we use 6 representing the pair on the x-axis. We find five significant cases for the simulated Gaussian map. We find 7, 14, 52 significant cases corresponding to simulated non-Gaussian cases for $J=1,2,3$ respectively. Finally, we show $U$-values on the y-axis obtained using the variable of Type $(vi)$ for simulated Gaussian and non-Gaussian maps in the Figure (\ref{fig:sim-t6}). Here again, we use the lower member of the $\ell$-mode pair to represent the pair on the x-axis. We find eight significant detections for the simulated Gaussian map. We find 10, 13, 53 significant cases for simulated non-Gaussian maps with $J=1,2,3$ respectively.

\subsection{Foreground contaminated 100 and 143 GHz Planck frequency Maps}
\label{result-1}
\begin{figure}[H]
		\vspace{-.5cm}
	\hspace{-1.2cm}
	\includegraphics[scale=0.58]{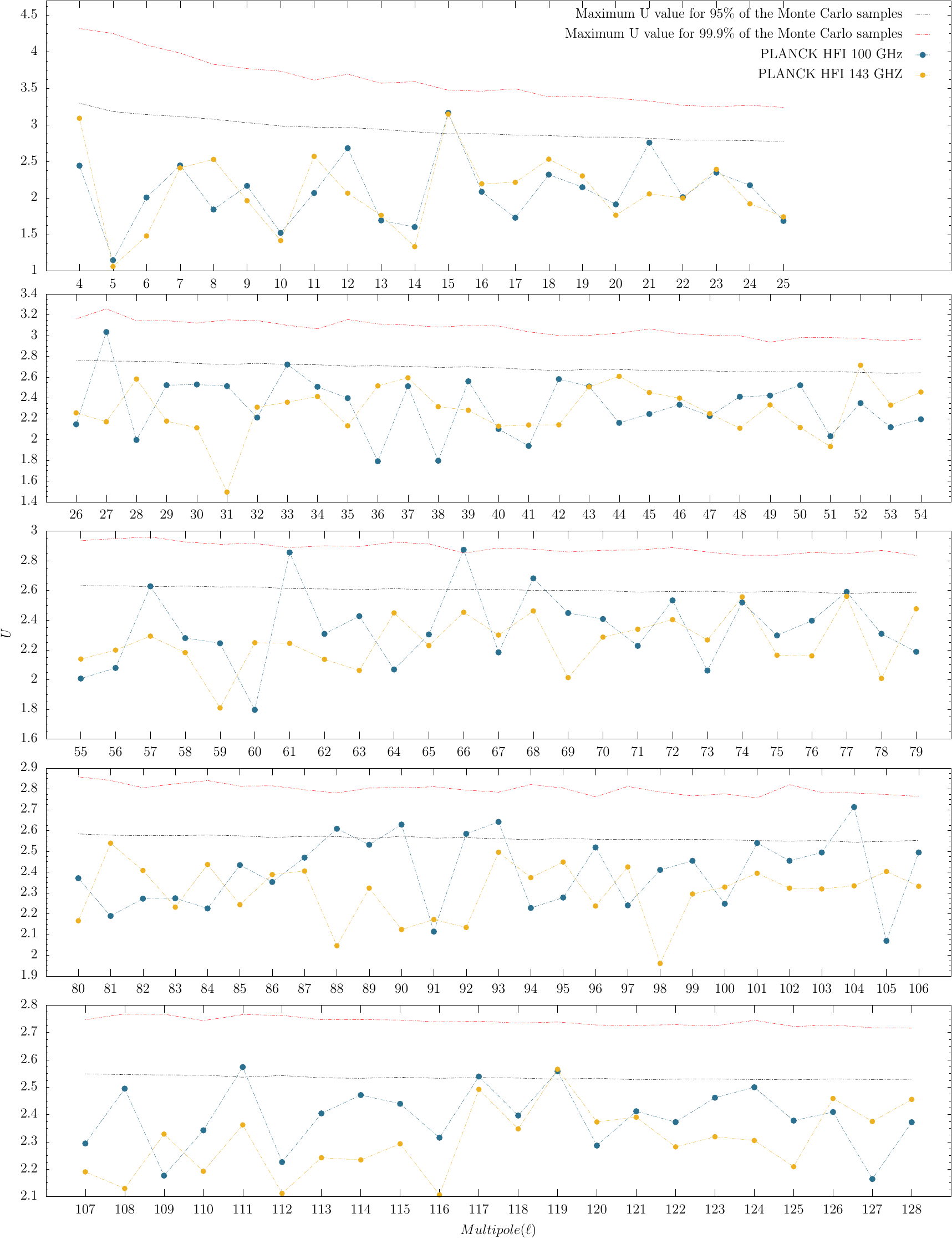}
	\caption{In the above figure we show $U$-values on vertical axis for testing uniformity of the given set of phases (variable of Type $(i)$) for various $\ell$-modes on horizontal axis for Planck 100 and 143 GHz foreground contaminated channel maps. We also show the maximum of $U$ for 95\% and 99.9\% of the corresponding Monte Carlo simulated samples.}
	\label{fig:fg-1}
\end{figure}

\begin{figure}[H]
		\vspace{-1cm}
	\hspace{-1.2cm}
	\includegraphics[scale=0.58]{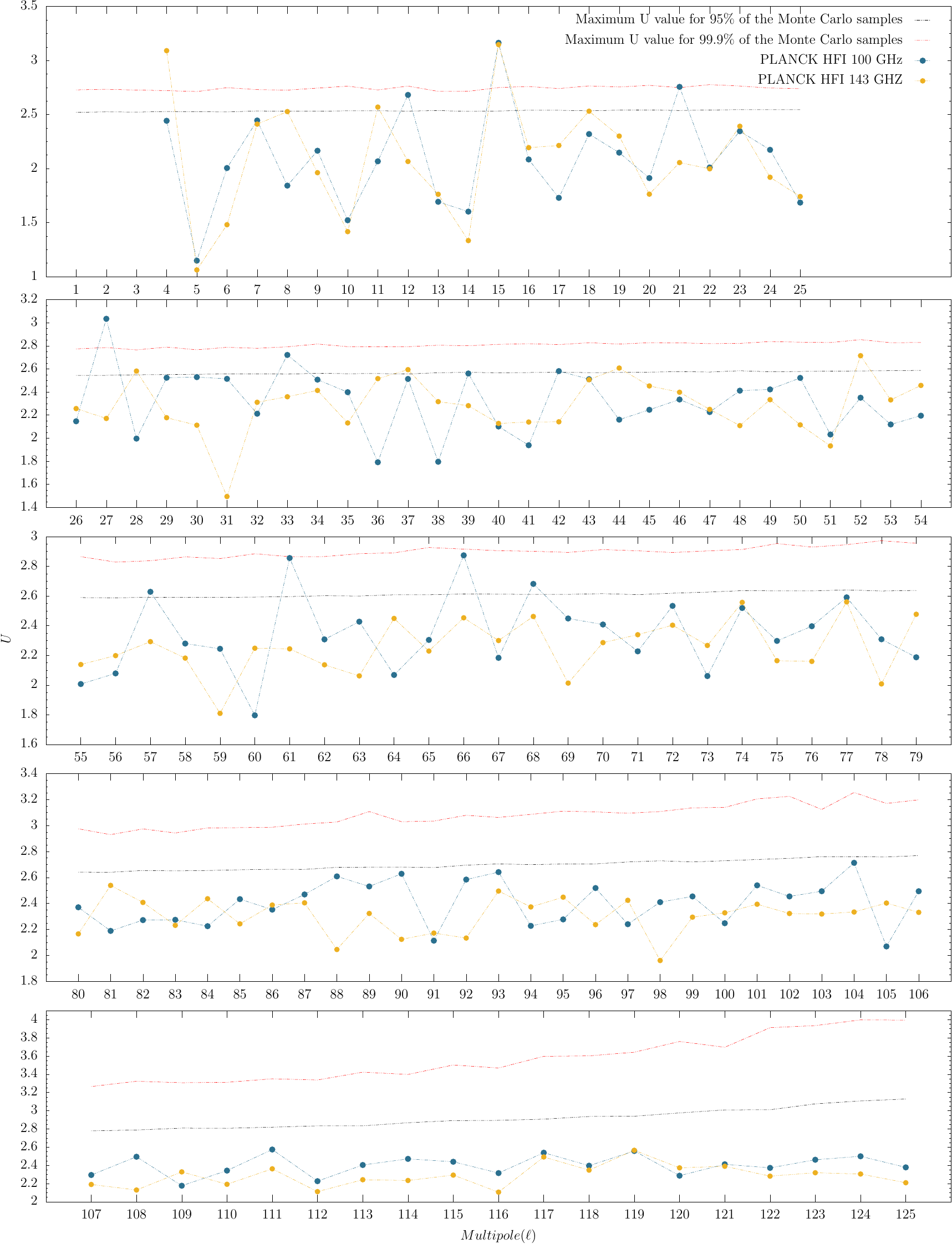}
	\caption{The figure shows the $U$-value on vertical axis and $m$ on horizontal axis for testing uniformity of SHP for given $m$ and all $\ell$ (variable of Type $(ii)$) for Planck 100 and 143 GHz foreground contaminated channel maps. We also show the maximum of $U$ for 95\% and 99.9\% of the corresponding Monte Carlo simulated phase samples.}
	\label{fig:fg-2}
\end{figure}

\begin{figure}[H]
	\vspace{-1cm}
	\hspace{-1.2cm}
	\includegraphics[scale=0.58]{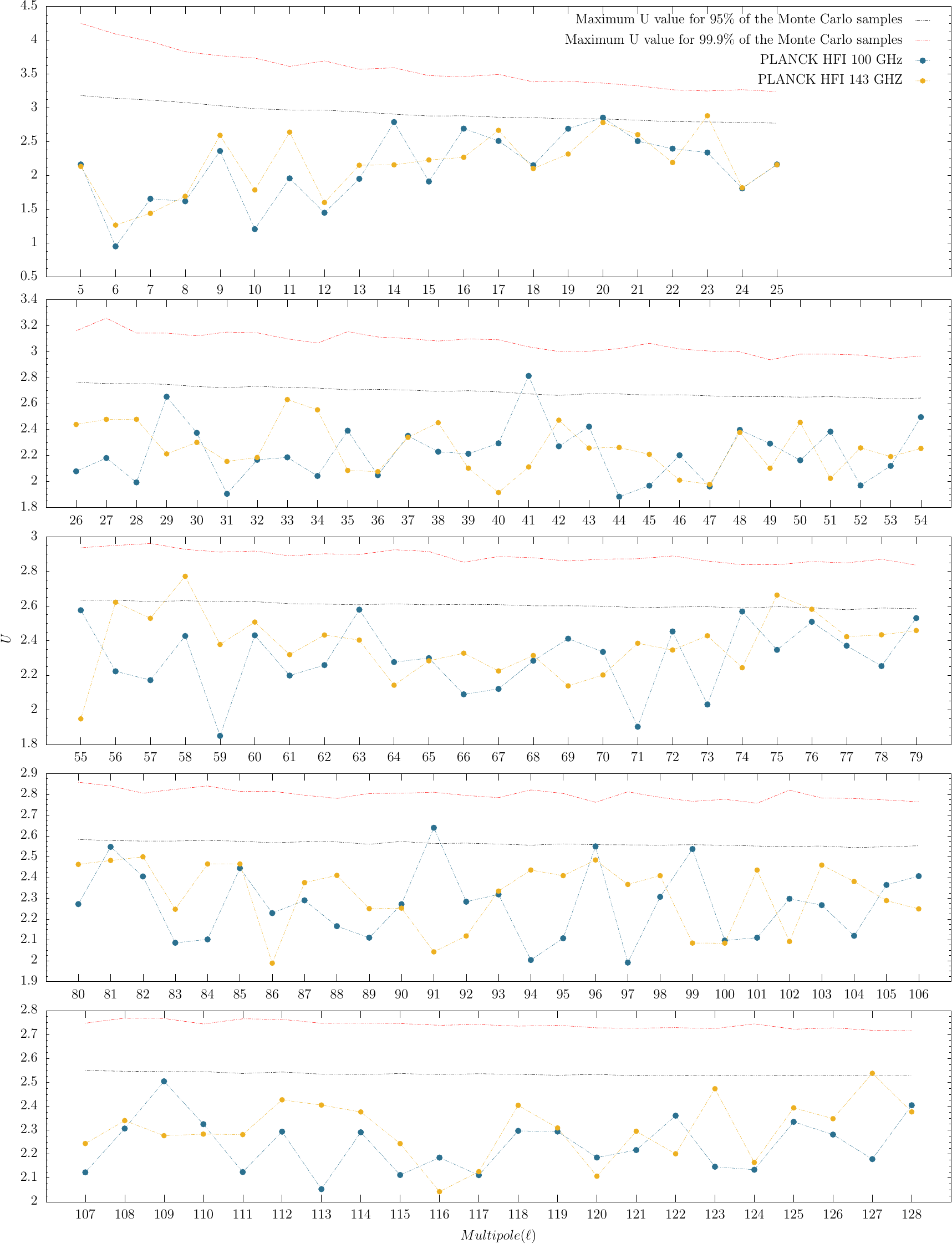}
	\caption{The above figure shows the $U$-value obtained for Planck 100 and 143 GHz foreground contaminated channel maps using variable of Type $(iii)$. The vertical and horizontal axes show $U$-value and $\ell$-modes respectively. We also show the maximum of $U$ for 95\% and 99.9\% of the corresponding Monte Carlo simulated uniform phases in $(0,2\pi]$.}
	\label{fig:fg-3}
\end{figure}

\begin{figure}[H]
	\vspace{-1.1cm}
	\hspace{-1.2cm}
	\includegraphics[scale=0.58]{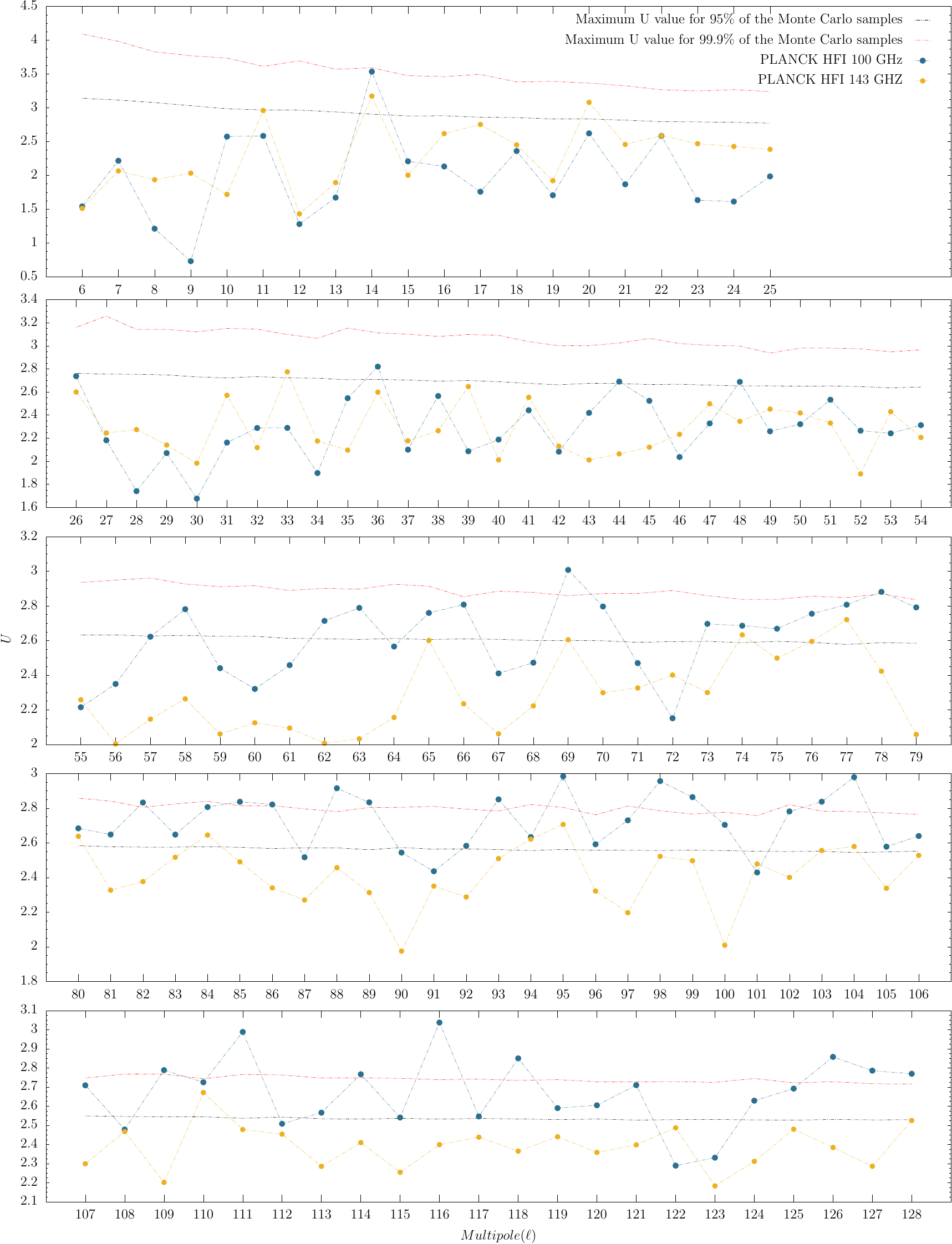}
	\caption{This figure shows $U$-value obtained using variable of Type $(iv)$ on phases for Planck 100 and 143 GHz foreground contaminated channel maps. The $\ell$-modes are shown on horizontal axis and $U$-values are shown on vertical axis. The maximum of $U$ for 95\% and 99.9\% of the corresponding simulated uniform phases in $(0,2\pi]$ are shown in black and red dashed lines respectively.}
	\label{fig:fg-4}
\end{figure}

\begin{figure}[H]	
	\vspace{-1cm}
	\hspace{-1.2cm}
	\includegraphics[scale=0.58]{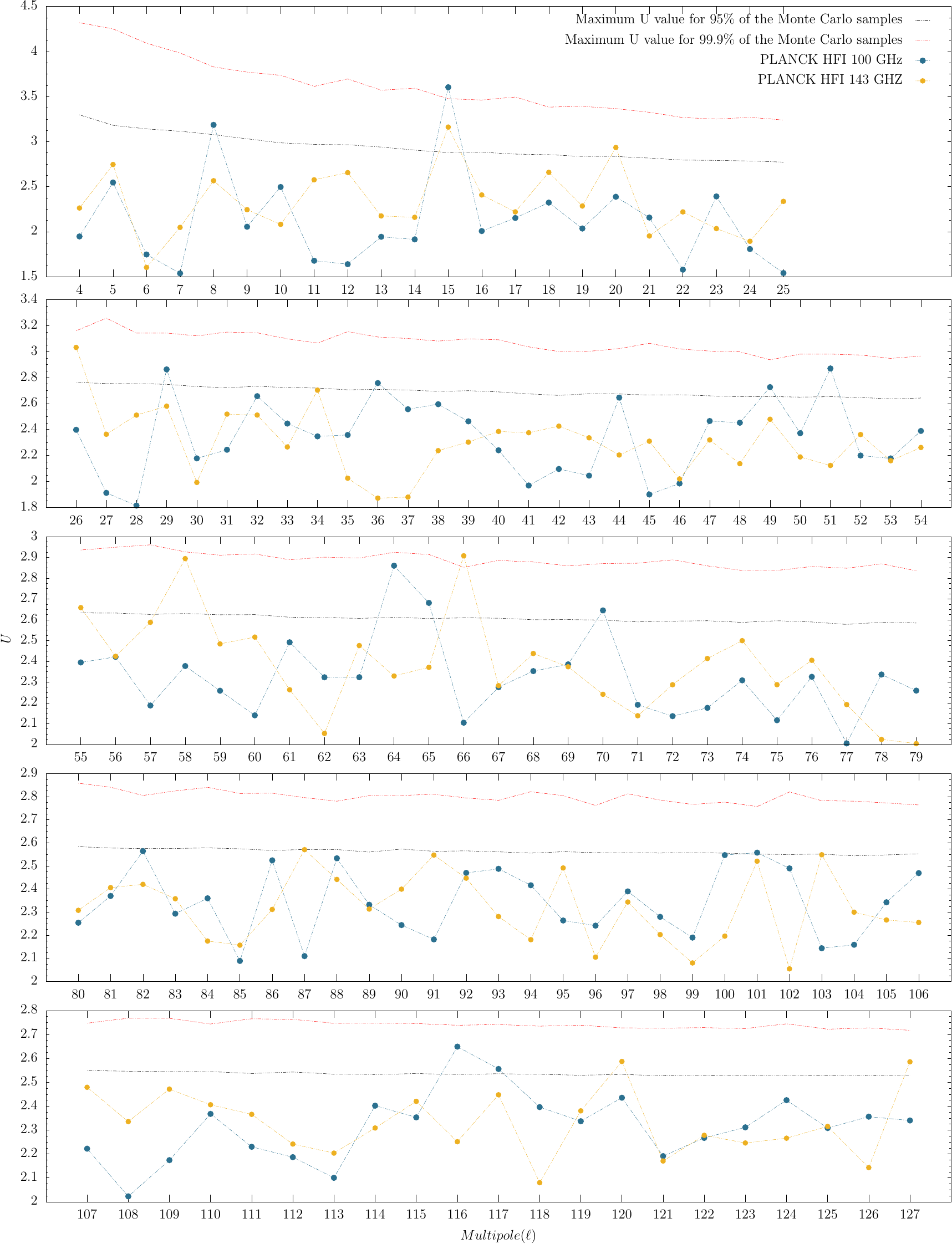}
	\caption{The figure shows $U$-value obtained using variable of Type $(v)$ for Planck 100 and 143 GHz foreground contaminated channel maps. The vertical and horizontal axes show $U$-value and lower of the $\ell$-mode pair respectively. We also show the maximum of $U$ for 95\% and 99.9\% of the corresponding Monte Carlo simulated uniform phase samples in $(0,2\pi]$ in black and red dashed lines respectively.}
	\label{fig:fg-5}
\end{figure}

\begin{figure}[H]	
	\vspace{-1cm}
	\hspace{-1.2cm}
	\includegraphics[scale=0.58]{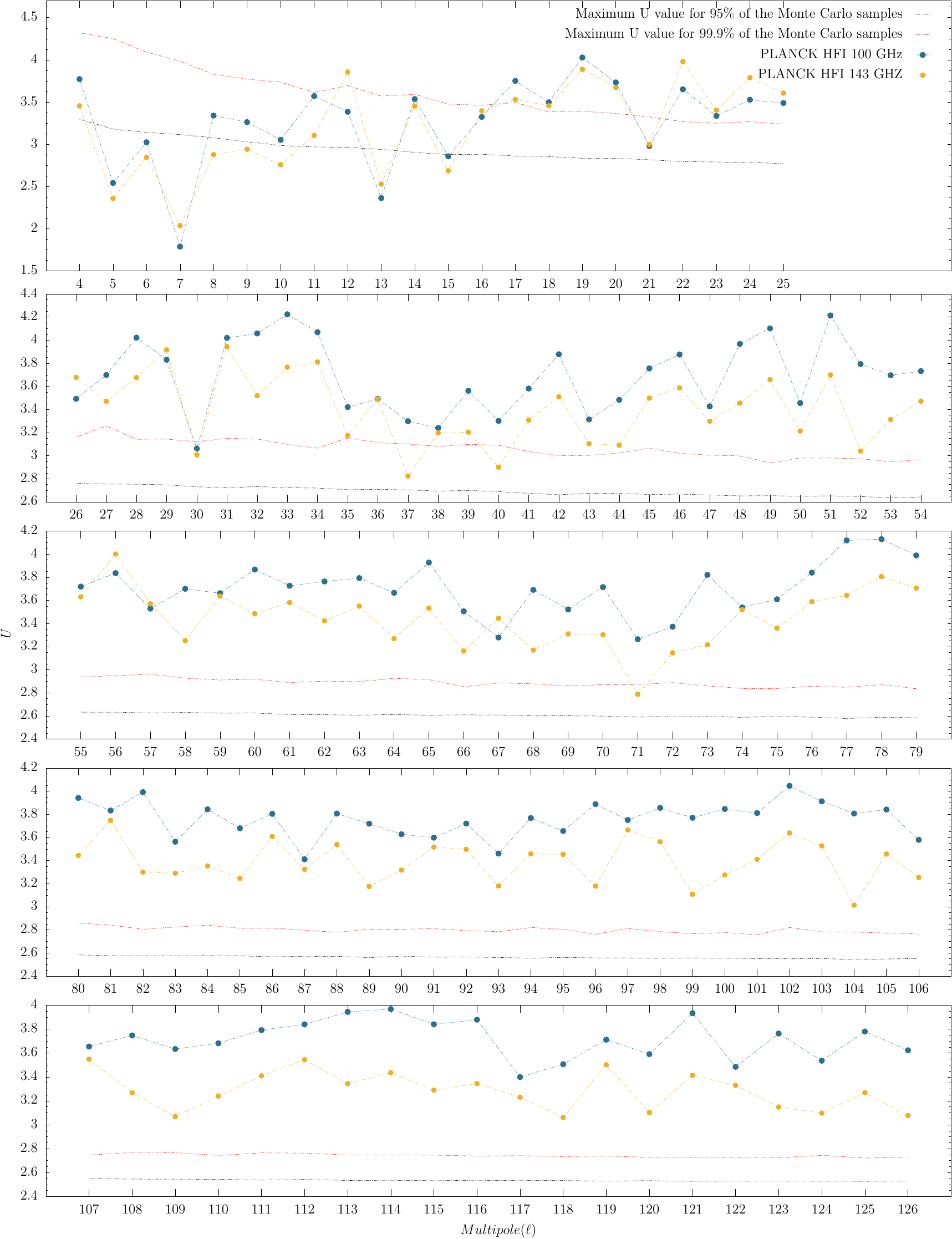}
	\caption{The figure shows $U$-value obtained using variable of Type $(vi)$ for Planck 100 and 143 GHz foreground contaminated channel maps. The vertical and horizontal axes show $U$-value and lower of the $\ell$-mode pair respectively. We also show the maximum of $U$ for 95\% and 99.9\% of the corresponding Monte Carlo simulated uniform phase samples in $(0,2\pi]$ in black and red dashed lines respectively.}
	\label{fig:fg-6}
\end{figure}

This subsection presents results obtained from the six tests applied on spherical harmonic phases of 100 GHz and 143 GHz Planck foreground contaminated channel maps. These maps can be considered as observed non-Gaussian maps, albeit without completely known statistical properties.
The results for these two foreground contaminated Planck channel maps are plotted in Figure (\ref{fig:fg-1}) to (\ref{fig:fg-6}).

In the Figure (\ref{fig:fg-1}) we plot $U$-values on y-axis obtained for Planck foreground contaminated 100 and 143 GHz channel maps for variable of Type $(i)$. On the plot, the x-axis represents the $\ell$-mode whose phases we test for uniformity. We find fifteen and three significant cases for 100 GHz, 143 GHz channel maps, respectively, against seven for the simulated Gaussian map. In the Figure (\ref{fig:fg-2}) we plot $U$-values on y-axis and $m$-modes on x-axis corresponding to variable of Type $(ii)$, for both the Planck channel maps. We find 63, 40 significant $m$-modes for 100 GHz, 143 GHz channel maps, respectively, against six for the simulated Gaussian map. We show $U$-values on y-axis obtained using variable of Type $(iii)$ in Figure (\ref{fig:fg-3}) for both 100 and 143 GHz channel maps. The x-axis shows the $\ell$-modes corresponding to which phase differences are used for the test. We find 3, 4 significant $m$-modes for 100 GHz, 143 GHz channel maps, respectively, against seven for the simulated Gaussian map. In Figure (\ref{fig:fg-4}) we show $U$-values on y-axis obtained for variable of Type $(iv)$ for 100 and 143 GHz Planck channel maps. In the plot, the x-axis represents the $\ell$-mode corresponding to which $m$-mode phase differences are used as test variables. We find 59, 13 significant $m$-modes for 100 GHz, 143 GHz channel maps, respectively, against 7 for the simulated Gaussian map. Figure (\ref{fig:fg-5}) shows $U$-values obtained after applying Rao's statistic on the variable of Type $(v)$ for 100 and 143 GHz channel maps. We show $U$-values on the y-axis and lower of the $\ell$-mode pair involved in the analysis on the x-axis. We find 12, 8 significant $\ell$-mode pairs for 100 GHz, 143 GHz channel maps, respectively, against five for the simulated Gaussian map. Finally, we show $U$-values on the y-axis obtained using the variable of Type $(vi)$ for 100 and 143 GHz Planck channel maps in the Figure (\ref{fig:fg-6}). We find 118, 115 significant $\ell$-mode pairs for 100 GHz, 143 GHz channel maps, respectively, against 8 for the simulated Gaussian map.




\subsection{Foreground Minimized CMB Maps}
\label{result-2}
\begin{figure}[H]
	\vspace{-1cm}
	\hspace{-1.2cm}
	\includegraphics[scale=0.58]{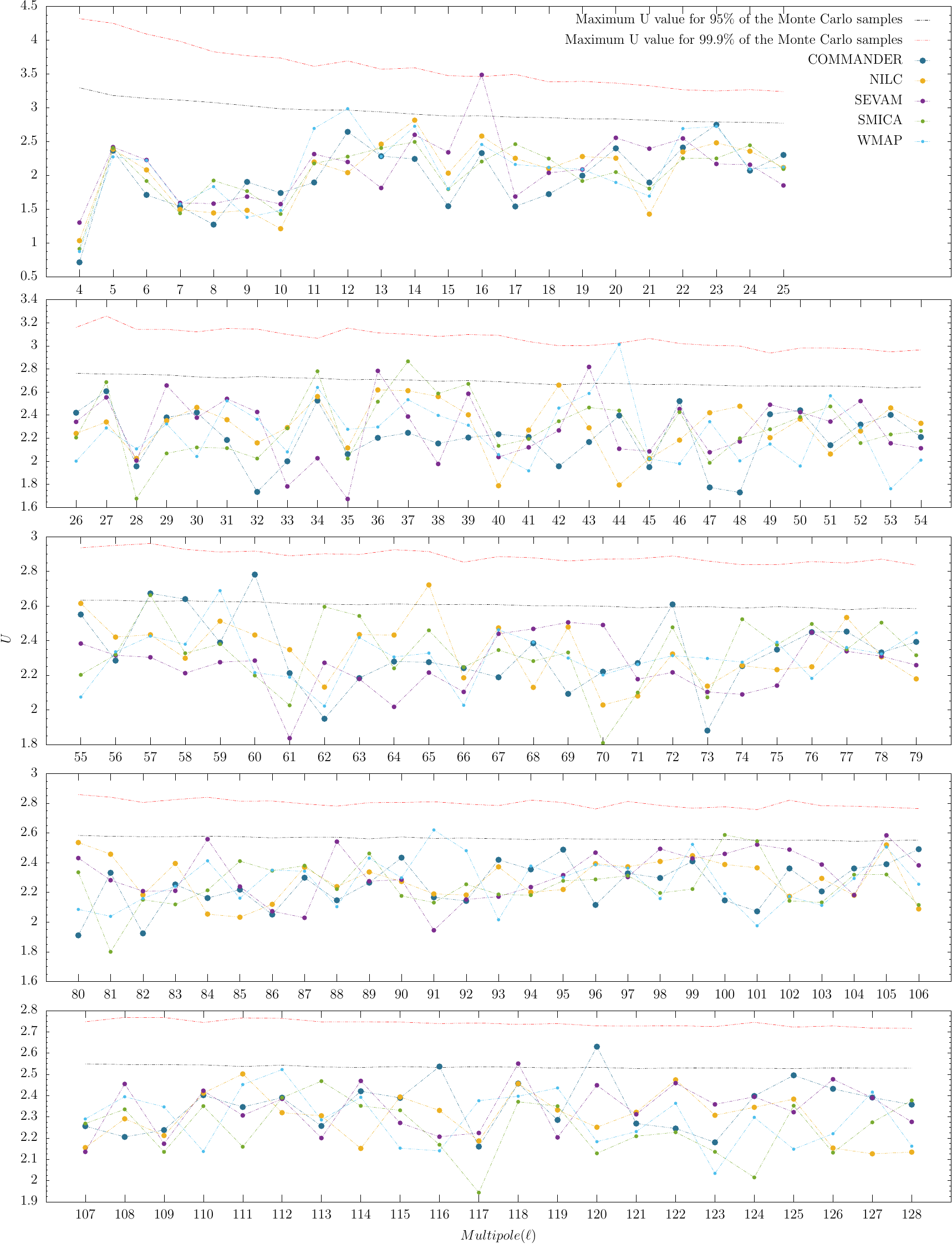}
	\caption{In the above figure we show $U$-values on vertical axis for testing uniformity of the given set of SHP (variable of Type $(i)$) for various $\ell$-modes on horizontal axis for all CMB cleaned maps. We also show the maximum of $U$ for 95\% and 99.9\% of the corresponding simulated samples.}
	\label{fig:cm-1}
\end{figure}

\begin{figure}[H]
	\vspace{-1cm}
	\hspace{-1.2cm}
	\includegraphics[scale=0.58]{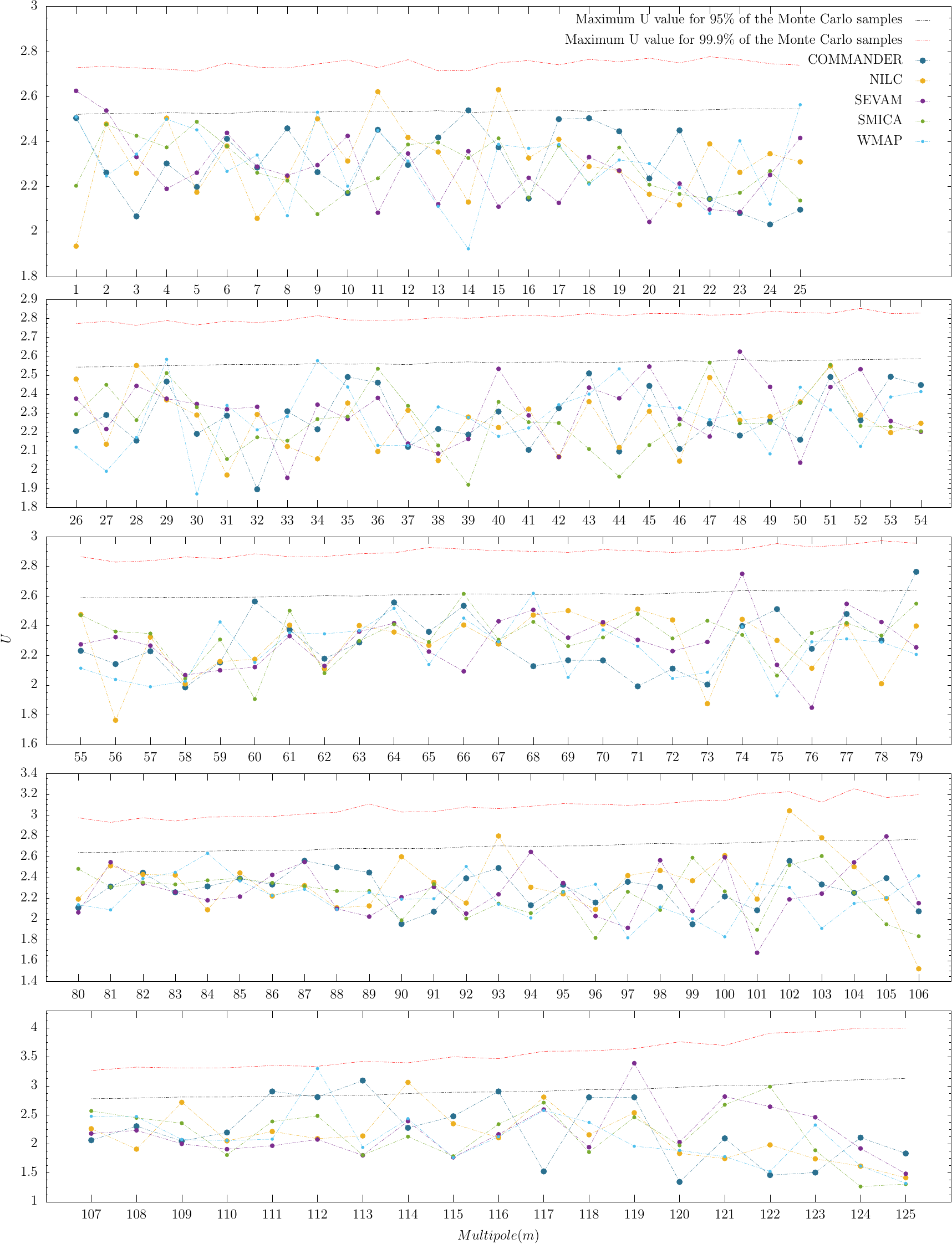}
	\caption{The figure shows the $U$-value on vertical and $m$-modes on horizontal axes for testing uniformity of SHP for given $m$ and all $\ell$ (variable of Type $(ii)$) for all CMB cleaned maps. We also show the maximum of $U$ for 95\% and 99.9\% of the corresponding Monte Carlo simulated phase samples.}
	\label{fig:cm-2}
\end{figure}

\begin{figure}[H]
	\vspace{-1cm}
	\hspace{-1.2cm}
	\includegraphics[scale=0.58]{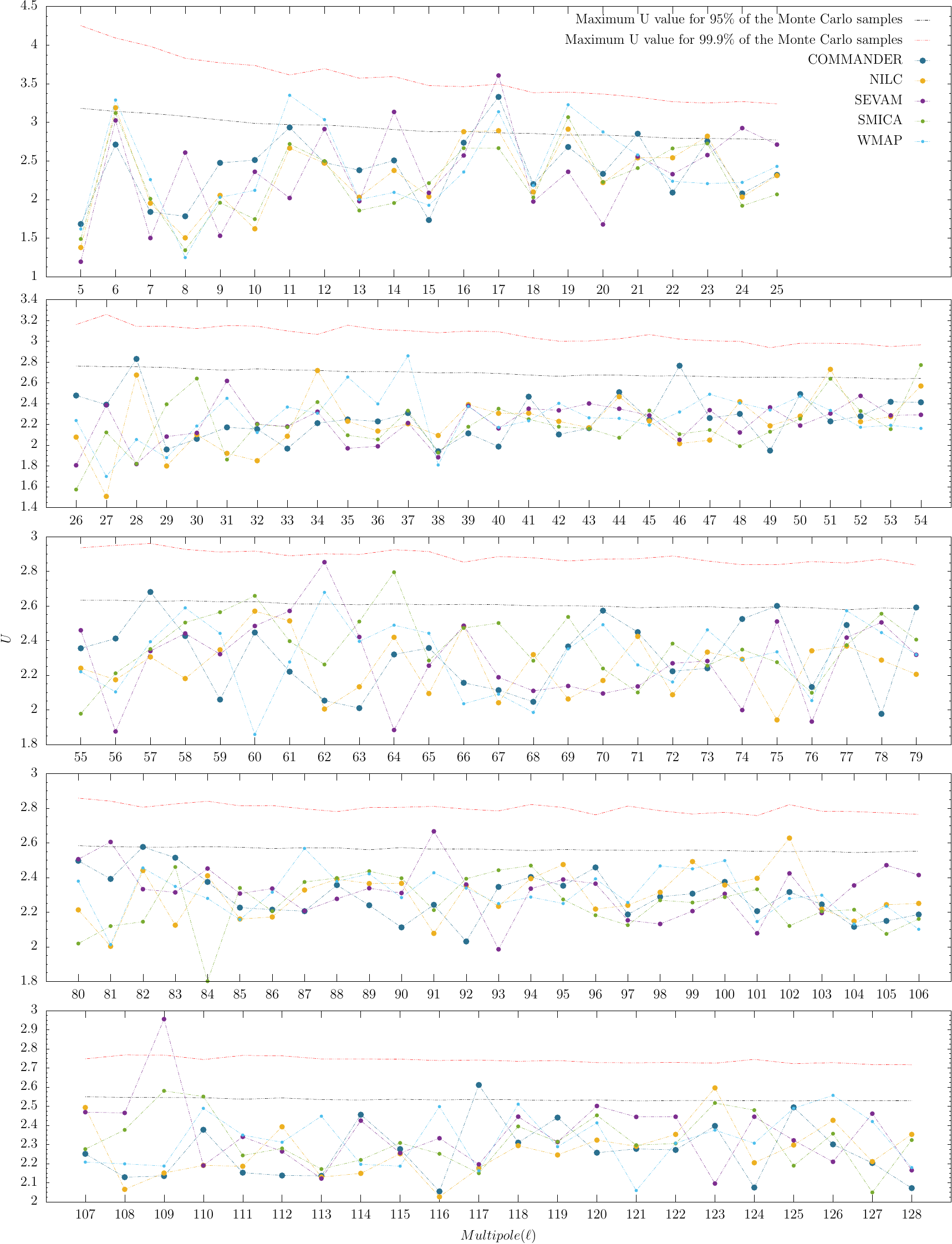}
	\caption{The above figure shows the $U$-value obtained for all CMB cleaned maps using variable of Type $(iii)$. The vertical and horizontal axes show $U$-value and $\ell$-modes respectively. We also show the maximum of $U$ for 95\% and 99.9\% of the corresponding Monte Carlo simulated uniform phases in $(0,2\pi]$.}
	\label{fig:cm-3}
\end{figure}

\begin{figure}
	\vspace{-1cm}
	\hspace{-1.2cm}
	\includegraphics[scale=0.58]{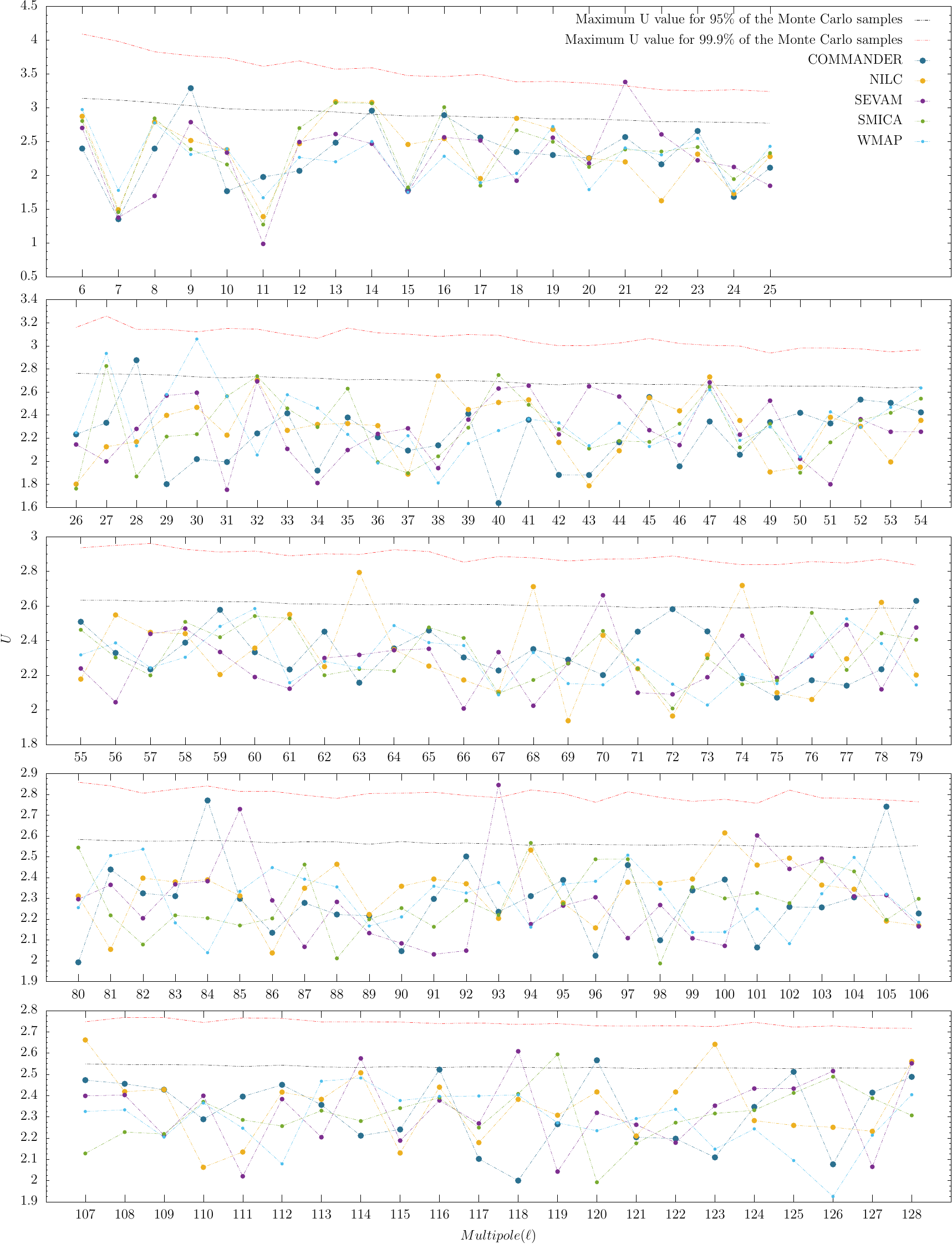}
	\caption{This figure shows $U$-value obtained using variable of Type $(iv)$ on phases for all CMB cleaned maps. The $\ell$-modes are shown on vertical axis and $U$-values are shown on y-axis. The maximum of $U$ for 95\% and 99.9\% of the corresponding Monte Carlo simulated uniform phases in $(0,2\pi]$ are shown in black and red dashed lines respectively.}
	\label{fig:cm-4}
\end{figure}

\begin{figure}
	\vspace{-.5cm}
	\hspace{-1.2cm}
	\includegraphics[scale=0.58]{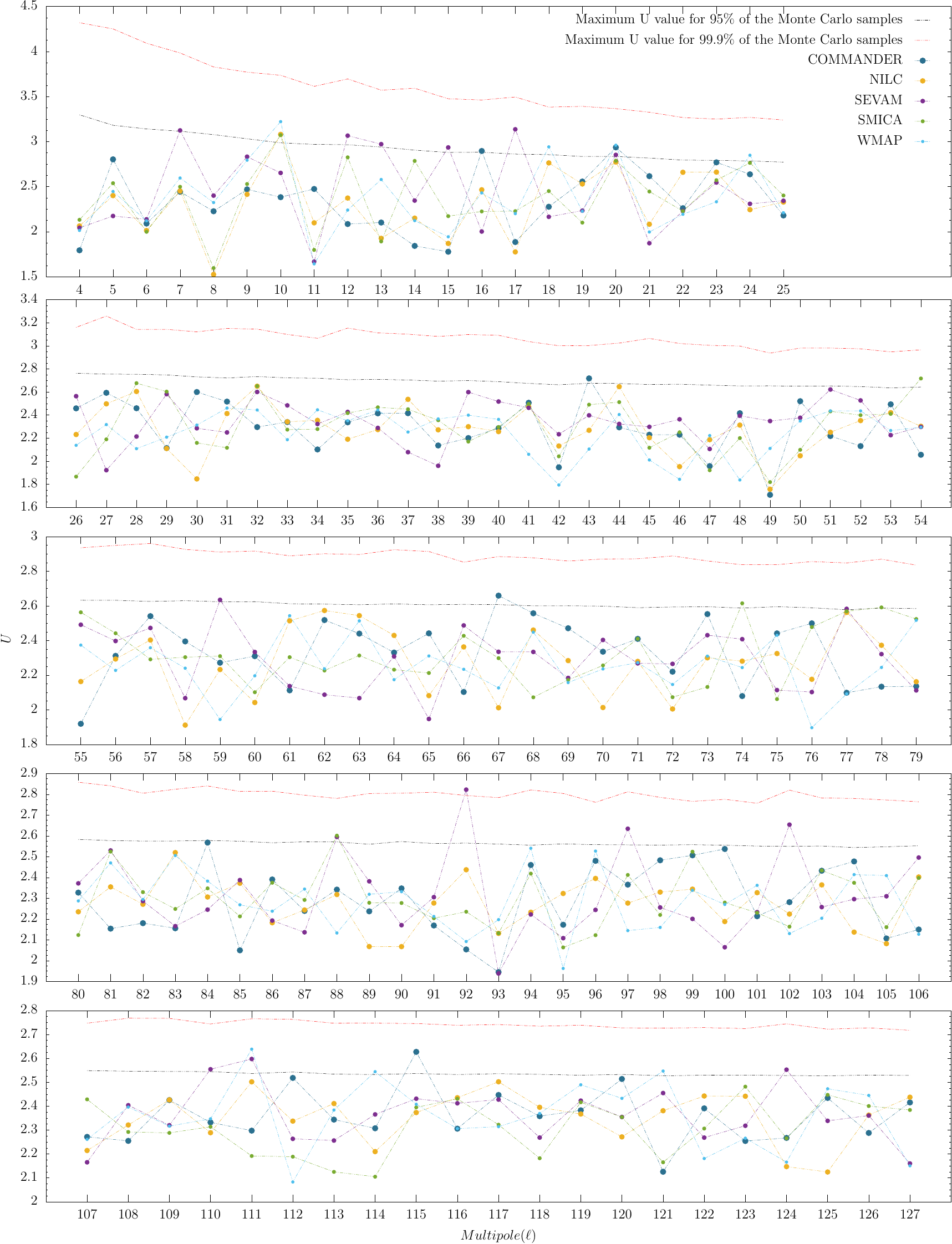}
	\caption{The figure shows $U$-value obtained using variable of Type $(v)$ for all CMB cleaned maps. The vertical and horizontal axes show $U$-value and lower of the $\ell$-mode pair respectively. We also show the maximum of $U$ for 95\% and 99.9\% of the corresponding Monte Carlo simulated uniform phase samples in $(0,2\pi]$ in black and red dashed lines respectively.}
	\label{fig:cm-5}
\end{figure}

\begin{figure}
		\vspace{-1cm}
	\hspace{-1.2cm}
	\includegraphics[scale=0.58]{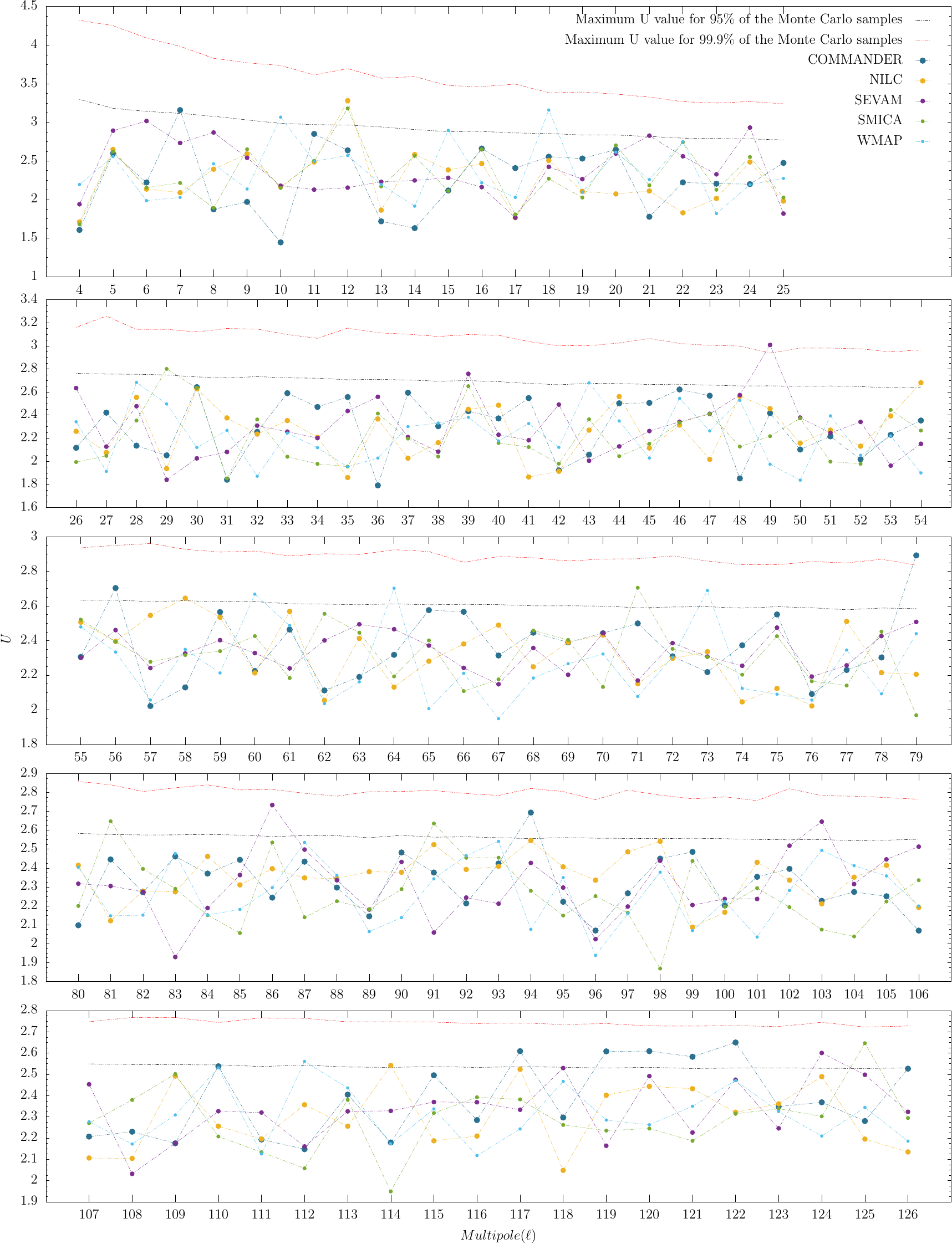}
	\caption{The figure shows $U$-value obtained using variable of Type $(vi)$ for all CMB cleaned maps. The vertical and horizontal axes show $U$-value and lower of the $\ell$-mode pair respectively. We also show the maximum of $U$ for 95\% and 99.9\% of the corresponding Monte Carlo simulated uniform phase samples in $(0,2\pi]$ in black and red dashed lines respectively.}
	\label{fig:cm-6}
\end{figure}

In this subsection we present results obtained from the six tests applied on spherical harmonic phases of COMMANDER, NILC, SEVEM, SMICA and WMAP ILC clean CMB temperature maps. The results for these five clean maps are plotted in Figure (\ref{fig:cm-1}) to (\ref{fig:cm-6}). We also show significant cases detected using each of the six tests, for all maps in, Table (\ref{table:cm-1}) to
(\ref{table:cm-6}). The first number shown in all of the Tables are the significant $\ell$ mode or representative $\ell$ mode, except in Table (\ref{table:cm-2}), where it represents the $m$-mode. The number in adjoining small bracket show P-values corresponding each of the significant detection.  

\vspace{1cm}  
    
Using the Type $(i)$ variables, we test for uniformity in phases of individual $\ell$-modes, the U-values for all the modes from $\ell=4$ to 128 are shown in the Figure (\ref{fig:cm-1}). Various statistically significant non-uniformity in phases using the Rao's statistic are summarized in Table (\ref{table:cm-1}) for COMMANDER, NILC, SEVEM, SMICA and WMAP ILC maps. For variable of Type (i) we find phase corresponding to $\ell$-mode 57, 58, 60, 72, 116 and 120 in COMMANDER are significant. For SMICA 34, 37, 57 and 100 $\ell$-modes are significant. We find $\ell$-mode 65 to be significant for NILC, 16, 36, 43, 105 and 118 are significant for SEVEM and 12, 44, 59, 91 for WMAP.

\vspace{1cm} 

Using the Type $(ii)$ variables, we test uniformity of SHP corresponding to all $\ell$-modes for a given $m$-mode. The U-values for all the modes from $\ell=4$ to 128 for this test are shown in the Figure (\ref{fig:cm-2}). The significant $m$-modes are shown in Table (\ref{table:cm-2}), along with their respective P-values. We find 14, 79 111 and and 113 in COMMANDER 66 in SMICA, 11, 15, 93, 102, 103 and 114 in NILC, 1, 2, 48, 74 105 and 119 in SEVEM and 25, 29, 34 and 112 in WMAP as significant $m$-modes. 

\vspace{1cm} 

Using Type $(iii)$ variables, we test for correlation among consecutive $m$ mode phases of a given $\ell$ mode. In the Figure (\ref{fig:cm-3}) corresponding to this case, multipole axis starts from $\ell=5$, as for Rao's statistic, we need at least four variables. The significant modes are shown in Table (\ref{table:cm-3}), along with their respective P-values. For the COMMANDER temperature maps the significant $\ell$-modes are 17, 21, 28, 46, 57, 75, 75, 79 AND 117. We find $\ell$-mode 19, 54, 60, 64, 109 and 110 for SMICA, 17, 19, 51, 102, 123 for NILC, 14, 17, 24, 62, 81, 91, 109 for SEVEM, 6, 11, 12, 17, 19, 20, 37, 62 and 126 for WMAP temperature maps as significant.

\vspace{1cm} 

Using Type $(iv)$ variables, we test for correlation, employing $U$ as test statistic, among next to consecutive $m$ mode phase of a given $\ell$ mode. The U-values for various multipoles for this case is shown in Figure (\ref{fig:cm-4}) and significant mutipoles and corresponding P-values is presented in Table (\ref{table:cm-4}). We find phases with statistically significant correlations between next to consecutive phase in $\ell$-modes 9, 14, 28, 79, 84, 105 and 120 for COMMANDER temperature map. We detect significant correlation in phase of next to consecutive phase in $\ell$ mode 13, 14, 16, 27, 40, 94 and 119 for SMICA, 13, 14, 38, 47, 63, 68, 74, 78, 100, 107, 123 and 128 for NILC, 21, 47, 70, 85, 93, 101, 114, 118 and 128 for SEVEM maps. For WMAP we found statistically significant detection at $\ell$ mode 27 and 30.

\vspace{1cm} 

Using type $(v)$ variables, we test for correlation of same $m$ mode phases of a given $\ell$ and it's consecutive $\ell$ mode. The U-values for this analysis are shown in the Figure (\ref{fig:cm-5}) with the statistically significant cases for all maps shown in Table (\ref{table:cm-5}). The multipole axis denotes the lower of the $\ell$ modes in the pair. We detect significant correlation between same $m$ mode phases of pair (16,17), (20,21), (43,44), (67,68), (115,116) for the COMMANDER map. For SMICA map, we find significant correlation between same $m$ mode phases of pair (10,11), (54,55), (74,75), (78,79) and (88,89). For NILC map, we find only pair (10,11) to be significant. The pair (7,8) (12,13), (13,14), (15,16), (7,18), (20,21), (59,60), (77,78), (88,89), (92,93), (97,98), (102,103), (110,111), (111,112), (124,125) are the significant occurrences for the SEVEM map. For WMAP ILC map we find pair (10,11), (18,19), (20,21), (24,25), (111,112), (114,115), (121,122) as significant. 

 \begin{center}
	\begin{tabular}{|m{1.8cm}|m{1.8cm}|m{1.8cm}|m{1.8cm}|m{1.8cm}|}
		\hline
		\textbf{COMM.}&\textbf{SMICA}&\textbf{NILC}&\textbf{SEVEM}&\textbf{WMAP}\\ 
		\hline 
		57(0.9682) 58(0.9541) 60(0.9916) 72(0.9574) 116(0.9532) 120(0.9892)&34(0.9690) 37(0.9878) 57(0.9638) 100(0.9693)&65(0.9862)&16(0.9991) 36(0.9709) 43(0.9880) 105(0.9682) 118(0.9602)& 12(0.9547) 44(0.9990) 59(0.9752) 91(0.9764) \\ 
		\hline																
	\end{tabular}
	\captionof{table}{ Table showing the significant and highly significant occurrences corresponding to Type $(i)$ variables for various clean CMB maps.}
	\label{table:cm-1}
\end{center}

Using Type $(vi)$ variables, we perform correlation test similar to that of Type $(v)$, but now we take difference of $m$ mode phases in $\ell$ and next to subsequent $\ell$. In Figure (\ref{fig:cm-6}) with lower of the $\ell$ in pair represented on the multipole axis, the p-values for various pair are plotted. We summaries the significant occurrences for this case and for all maps in Table (\ref{table:cm-6}). We find statistically significant detection in pair (7,9), (56,58), (79,81), (94,96), (117,119), (119,121), (120,122), (121,123), (122,124) for the COMMANDER map. We find pair (12,13), (29,31), (71, 73), (81,83), (91,93), (125,127) as statistically significant for the SMICA map. For the NILC map we find significant correlation between  $\ell$ mode pair (12,14), (54,56), (58,60), (114,116). For the SEVEM map, the statistically significant pairs are (21,23), (24,26), (39,41), (49,51), (86,88), (103,105), (124,126). For the WMAP, we find the pair (10,12), (15,17), (18,20), (43,45), (60,62), (64,66), (73,75) and (112,114) as statistically significant correlated.

	\begin{center}
	\begin{tabular}{|m{1.8cm}|m{1.8cm}|m{1.8cm}|m{1.8cm}|m{1.8cm}|}
	\hline
	\textbf{COMM.}&\textbf{SMICA}&\textbf{NILC}&\textbf{SEVEM}&\textbf{WMAP}\\ 
	\hline 
	14(0.9522) 79(0.9842) 111(0.9649) 113(0.9874) &  66(0.9520)&11(0.9872) 15(0.9882) 93(0.9773) 102(0.9946) 103(0.9572) 114(0.9790)&1(0.9914)  2(0.9584) 48(0.9727) 74(0.9854) 105(0.9556) 119(0.9920)& 25(0.9631) 29(0.9654) 34(0.9586) 112(0.9969)   \\ 
	\hline																
\end{tabular}
	\captionof{table}{ Table showing the significant and highly significant occurrences corresponding to Type $(ii)$ variables for various clean CMB maps.}
	\label{table:cm-2}
\end{center}

	\begin{center}
		\begin{tabular}{|m{1.8cm}|m{1.8cm}|m{1.8cm}|m{1.8cm}|m{1.8cm}|}
	\hline
	\textbf{COMM.}&\textbf{SMICA}&\textbf{NILC}&\textbf{SEVEM}&\textbf{WMAP}\\ 
	\hline 
	17(0.9975) 21(0.9604) 28(0.9697) 46(0.9813) 57(0.9692) 75(0.9553) 79(0.9555) 117(0.9827)&19(0.9874) 54(0.9875) 60(0.9634) 64(0.9946) 109(0.9699) 110(0.9565)&17(0.9541) 19(0.9652) 51(0.9738) 102(0.9817) 123(0.9829)& 14(0.9827) 17(0.9999) 24(0.9793) 62(0.9978) 81(0.9635) 91(0.9859) 109(1.000)&6(0.9615) 11(0.9922) 12(0.9630) 17(0.9885) 19(0.9960) 20(0.9605) 37(0.9880) 62(0.9761) 126(0.9672)\\ 
	\hline 
\end{tabular}
		\captionof{table}{Table showing the significant and highly significant occurrences corresponding to Type $(iii)$ variables for various clean CMB maps.}
		\label{table:cm-3}
	\end{center}

\begin{center}

	\begin{tabular}{|m{1.8cm}|m{1.8cm}|m{1.8cm}|m{1.8cm}|m{1.8cm}|}
	\hline
	\textbf{COMM.}&\textbf{SMICA}&\textbf{NILC}&\textbf{SEVEM}&\textbf{WMAP}\\ 
	\hline 
	9(0.9753) 14(0.9537) 28(0.9769) 79(0.9702) 84(0.9973) 105(0.9980) 120(0.9686)&13(0.9667) 14(0.9720) 16(0.9686) 27(0.9684) 40(0.9667) 94(0.9544) 119(0.9837)&13(0.9691) 14(0.9729) 38(0.9583) 47(0.9741) 63(0.9947) 68(0.9859) 74(0.9909) 78(0.9658) 100(0.9767) 107(0.9925) 123(0.9920) 128(0.9699)&21(0.9992) 47(0.9557) 70(0.9726) 85(0.9933) 93(0.9995) 101(0.9755) 114(0.9727) 118(0.9852) 128(0.9650)&27(0.9862) 30(0.9959)\\ 
	\hline 
\end{tabular}
	\captionof{table}{Table showing the significant and highly significant occurrences corresponding to Type $(iv)$ variables for various clean CMB maps.}
	\label{table:cm-4}
	
\end{center}

\begin{center}
	\begin{tabular}{|m{1.8cm}|m{1.8cm}|m{1.8cm}|m{1.8cm}|m{1.8cm}|}
	\hline
	\textbf{COMM.}&\textbf{SMICA}&\textbf{NILC}&\textbf{SEVEM}&\textbf{WMAP}\\ 
	\hline 
	16(0.9540) 20(0.9729) 43(0.9667) 67(0.9722) 115(0.9874)&10(0.9644) 54(0.9761) 74(0.9643) 78(0.9522) 88(0.9670)&10(0.9654)&   7(0.9518) 12(0.9686) 13(0.9562) 15(0.9620) 17(0.9894) 20(0.9541) 59(0.9554) 77(0.9536) 88(0.9629) 92(0.9996) 97(0.9830) 102(0.9887) 110(0.9568) 111(0.9800) 124(0.9651)& 10(0.9810) 18(0.9707) 20(0.9772) 24(0.9676) 111(0.9898) 114(0.9577) 121(0.9634)\\ 
	\hline 
\end{tabular}
	\captionof{table}{Table showing the significant occurrences corresponding to Type $(v)$ variables for various clean CMB maps. Multipoles represent the lower of the pair of modes involved in the test.}
	\label{table:cm-5}
\end{center}

 \begin{center}
 	
 	\begin{tabular}{|m{1.8cm}|m{1.8cm}|m{1.8cm}|m{1.8cm}|m{1.8cm}|}
	\hline
	\textbf{COMM.}&\textbf{SMICA}&\textbf{NILC}&\textbf{SEVEM}&\textbf{WMAP}\\ 
	\hline 
	7(0.9577)
	56(0.9767)
	79(0.9997)
	94(0.9924)
	117(0.9838)
	119(0.9877)
	120(0.9850)
	121(0.9789)
	122(0.9944)&      12(0.9818)
	29(0.9648)
	71(0.9882)
	81(0.9797)
	91(0.9820)
	125(0.9926)&      12(0.9890)
	54(0.9643)
	58(0.9564)
	114(0.9563) &      21(0.9522)
	24(0.9808)
	39(0.9701)
	49(0.9998)
	86(0.9947)
	103(0.9872)
	124(0.9856)&      10(0.9639)
	15(0.9543)
	18(0.9918)
	43(0.9516)
	60(0.9697)
	64(0.9805)
	73(0.9841)
	112(0.9615)      
	\\ 
	\hline 
\end{tabular}
 	\captionof{table}{Table showing the significant and highly significant occurrences corresponding to Type $(vi)$ variables for various clean CMB maps.  Multipoles represent the lower of the pair of modes involved in the test.}
 	\label{table:cm-6}
 	\end{center}	

\section{Discussions and Conclusion}
\label{Diss}
In this work we have used Rao's statistic to test the "random phase hypothesis" \citep{1986ApJ...304...15B,2013rossmanith}, which is our null hypothesis for a Gaussian random field, by investigating phase correlation between modes for various observed CMB maps. Any violation of either uniformity or independence in CMB SHP indicates potential signatures of non-Gaussianity in the corresponding spherical harmonic coefficients. The investigation of the random phase hypothesis involving spherical harmonic phase analysis of CMB maps has been rather~ limited in literature~\cite{2004MNRAS.350..989C,2006IJMPD..15.1283C,2013AN....334.1020K, 2005astro.ph..5046L, 2003ApJ...590L..65C, 2004ApJ...602L...1C, 2004MNRAS.349..695N, 2005PhRvD..72f3512N, 2004astro.ph..1622W}. In this work, we substantially and 
significantly extended the earlier results by employing a novel model-independent method using Rao's statistic for investigating non-Gaussian signatures in CMB temperature maps by testing random phase hypothesis to signal dominated scales limited by $\ell \le 128 $. The method looks into uniformity and correlations within individual $\ell$ and $m$ mode, and among nearby $\ell$ mode SHP, where any non-uniformity or correlation detected indicates violation of random phase hypothesis thereby pointing to non-Gaussianity. We found phases corresponding to several $\ell$, $m$ multipoles and pairs of $\ell$ multipoles which are inconsistent with Gaussian CMB shown in Tables (\ref{table:cm-1}) to (\ref{table:cm-6}).


The CMB component reconstructed maps (COMMANDER, SMICA, NILC, SEVEM, WMAP-ILC) discussed in this work have been obtained by various science groups employing independent statistical techniques by removing foreground emissions. This causes the morphology and hence phases obtained from these CMB maps to be some what different, although all these CMB maps represent the same last scattering surface. 
Any significant detection found across different maps is more likely to be of cosmological origin than the ones in a single map, provided the possibility of them being originating from other sources, and unaccounted systematic are ruled out. Not only any detected non-Gaussianity in the CMB temperature maps are essential to understand the distribution of primordial perturbations, but also these detected non-Gaussianity have the potentials to constrain residual systematic.

For comprehensive testing of the random phase hypothesis, we perform our analysis using three different classes of test variables. In class I, we test uniformity of phases for constant $\ell$ and $m$ modes, represented by Type $(i)$ and $(ii)$ 
test respectively. In class II, we test the correlation between consecutive and next to consecutive $m$ modes for a given $\ell$, represented by test Type $(iii)$ and $(iv)$ respectively. Finally, in class III, we test the correlation between consecutive and next to consecutive $\ell$ modes but same $m$-mode, represented by Type $(v)$ and $(vi)$ tests, respectively.

In order to test the performance of our method, we generated non-Gaussian CMB simulations using $a_{\ell m}$ samples from increasingly non-Gaussian distributions. We generated three different sets of non-Gaussian maps with varying non-Gaussianity. To test the performance of our method we applied all six test on 5000 different Monte Carlo simuated non-Gaussian maps of each set. From the Figure (\ref{fig:effic}), we find that number of significant detections increases as we apply our method on maps with increasing non-Gaussianity for all the tests as expected. Also, the  small error bars indicates that our method is very efficient in detecting non-Gaussianinity.

Apart from doing a study on ensemble of non-Gaussian maps, we also evaluate the efficiency of our method by randomly choosing a specific realization of Gaussian and non-Gaussian CMB maps. We summarize the result obtained by applying the test on the non-Gaussian maps against a simulated Gaussian map in Table (\ref{table:dc-1}). From the Table (\ref{table:dc-1}), we find that all the six tests detect more number of significant detection with increasing non-Gaussianity than the Gaussian CMB maps except Type (iv) test for the least non-Gaussian case.

 \begin{center}
	
	\begin{tabular}{|c|c|c|c|c|c|c|}
		\hline
		Modes above 95\% &Type $(i)$&Type $(ii)$ &Type $(iii)$&Type $(iv)$&Type $(v)$&Type $(vi)$\\ 
		\hline 
		$N_{NG1}/N_{G}$ &23/7  &23/6  &9/7  &5/8  &7/5  &10/8\\ 
		$N_{NG2}/N_{G}$ &69/7  &74/6  &15/7 &13/8  &14/5  &13/8\\ 
		$N_{NG3}/N_{G}$ &116/7 &116/6 &67/7 &51/8 &52/5 &53/8\\
		\hline 
	\end{tabular}
	\captionof{table}{The table summarizes the number of significant cases for four simulated non-Gaussian maps $N_{NG}$ against the number of significant cases for simulated Gaussian map $N_{G}$ for a test involving each of the variable Type.}
	\label{table:dc-1}
\end{center}  

Next we apply our method on Planck foreground contaminated HFI 100 and 143 GHz temperature maps. These maps can be considered as observed non-Gaussian maps for which complete statistical properties are not known due to presence of foregrounds (or any possible unknown systematic).This provides a case for the application of our method on data and serves as an alternative test of the performance of our estimator since these maps are expected to contain non-Gaussian signals. The fact that our method detects a significant non-Gaussian signal in these maps supports that it can detect a non-Gaussian signal different from what has been included in the simulations. 
	 We summarize the result obtained applying the method on foreground contaminated Planck 100 and 143 GHz HFI maps in Table (\ref{table:dc-2}). From the Tables (\ref{table:dc-2}) and (\ref{table:dc-1}) together, we find that all six tests discussed in section (\ref{method}) are useful in detecting phase correlations.     
\begin{center}
	
	\begin{tabular}{|c|c|c|c|c|c|c|}
		\hline
		Modes above 95\% &Type $(i)$&Type $(ii)$ &Type $(iii)$&Type $(iv)$&Type $(v)$&Type $(vi)$\\ 
		\hline 
		$N_{100}/N_{G}$&15/7&63/6 &3/7 &59/8 &12/5 &118/8\\ 
		$N_{143}/N_{G}$&3/7&40/6 &4/7 &13/8 &8/5 &115/8\\ 
		
		\hline 
	\end{tabular}
	\captionof{table}{The table summarizes the number of significant cases for Planck HFI 100 GHz $N_{100}$ and 143 GHz $N_{143}$, against the number of significant cases for simulated Gaussian maps $N_{G}$, for a test involving each of the variable Type.}
	\label{table:dc-2}  
\end{center}

Finally we apply our method on foreground cleaned CMB maps. Applying the method to test the uniformity of SHP for a given $m$-mode across $\ell$, (type (ii) variables) we find that most of the $m$-mode phases for a given $m$ are uniform. We do not find any common significant $m$-mode phases across maps. Modes 79, 113 in COMMANDER, 11, 15, 102 in NILC 1, 74, 119 in SEVEM and 112 in WMAP are possible $m$-mode phases for further investigation, given their relatively high P-values.  

We perform our analysis into three different parts together for test variable type (i), (iii) and (iv) which are designed to detect correlations of phases within a 
given $\ell$ mode. We look for statistically significant multiple occurrences for the same type of variables but different maps, different types of variables but same map and all type of variables all maps for test variable Type (i), (iii) and (iv). With this analysis method, we find that, for the type ($i$) variables, phases corresponding to most of the modes are uniform, with exceptions for $\ell$ mode 57 which occur in COMMANDER, SMICA. When considering high significance cases only, thereby decreasing the probability of Type I statistical error, we conclude that the phases corresponding to SMICA, NILC are all uniform. In contrast, some mode phases corresponding to COMMANDER, SEVEM, and WMAP are still non-uniform. For the type ($iii$) variables, phases corresponding to $\ell$ mode 17, 19, 62 and 109 are significant multiple times across cleaned maps, indicating there might be some interesting signal. For the variables of type ($iv$), phases corresponding to $\ell$ mode 13, 14, 27, 47 and 128 is statistically significant many times across the maps. Looking into the statistical significance across variables of type ($i$), ($iii$), ($iv$) for a given map, we find that for the COMMANDER map, $\ell$ modes 14, 57, 28, and 79 are significant more than one time. Similar repetition of significance for $\ell$ mode 123 for NILC map, 118 for SEVEM map, and 12 for WMAP are found. We do not find any such repetition for the SMICA map. Considering occurrences of the significance of an $\ell$ mode across the maps and across the variables, we conclude that $\ell$ mode 57 is statistically significant multiple times. To summarise phases of $\ell$ mode , 12, 13, 14, 17, 19, 27, 28, 47, 57, 62, 9, 118, 119, 123 and 128 have significance occurring multiple times in violation of random phase hypothesis indicating presence of non-Gaussian signature in corresponding spherical harmonic coefficients which might be of primordial origin.  

For the class III tests where we investigate the correlation between neighboring  $\ell$ modes, we look for statistically significant multiple occurrences of correlated pairs. For the variable type ($v$), where we investigate the correlation between subsequent $\ell$ modes, we find $\ell$ mode pairs (10,11), (20,21), (88,89) and (111,112) occur multiple times with significant correlations across cleaned maps. We find mode pair (20,21) in COMMANDER and SEVEM  maps. For the class II and variable type ($v$), where we investigate the correlation between $\ell$ and next to subsequent $\ell$ mode phases, we find that the mode pairs (12,14) is significant multiple times across different cleaned maps. The presence 
of non-uniform or correlated spherical harmonic phases and correlated phases corresponding to various mode pairs in the cleaned CMB maps indicate presence of non-Gaussian signals therein. An important future project will be to  investigate the underlying cause of the detected non-Gaussianities of this work.

\acknowledgments
SKY acknowledges financial support from Ministry of Human Resource and Development, Government of India via Institute Fellowship at IISER Bhopal, 
during the course of this work. The authors also thank the anonymous referee for her/his insightful comments which lead to useful improvement to the original article. We also thank  \href{https://www.pstat.ucsb.edu/people/s-rao-jammalamadaka}{\nolinkurl{S.} {Rao} {Jammalamadaka}} for fruitful discussions on Rao's statistics. We use the publicly available HEALPix~(\cite{2005ApJ...622..759G}) package to perform spherical harmonic decomposition and for visualization purposes from~\url{http://healpix.sourceforge.net}. We acknowledge the use of the Legacy Archive for Microwave Background Data Analysis (LAMBDA) and Planck Legacy Archive (PLA). LAMBDA is a part of the High Energy Astrophysics Science Archive Center (HEASARC) and the Planck Legacy Archive (PLA) contains all public products originating from the Planck mission, an ESA science mission with instruments and contributions directly funded by ESA Member States, NASA, and Canada. This research has made use of NASA's Astrophysics Data System.



\providecommand{\href}[2]{#2}\begingroup\raggedright\endgroup

\end{document}